\documentclass[12pt,onecolumn] {article} 
\usepackage{amsmath}
\usepackage{amssymb}
\usepackage{latexsym}
\usepackage{graphicx}

\begin{document}

\begin{center}
\baselineskip=24pt

{\Large \bf First measurement of low intensity fast neutron background from rock 
at the Boulby Underground Laboratory}
\vspace{0.5cm}

{\large
E.~Tziaferi~\footnote{Corresponding author, E-mail address: e.tziaferi@sheffield.ac.uk},
M.~J.~Carson,
V.~A.~Kudryavtsev,
R.~Lerner,
P.~K.~Lightfoot, 
S.~M.~Paling,
M.~Robinson, 
N.~J.~C.~Spooner}

\vspace{0.5cm}

{\it Department of Physics and Astronomy, 
University of Sheffield, Sheffield S3 7RH, UK}

\date{Today's Date}

\vspace{0.5cm}
\begin{abstract}

A technique to measure low intensity fast neutron flux has been developed. 
The design, calibrations, procedure for data analysis and interpretation 
of the results are discussed in detail. 
The technique has been applied to measure the 
neutron background from rock at the Boulby Underground Laboratory, 
a site used for dark matter and other experiments, requiring shielding from
cosmic ray muons. The experiment was performed using a liquid
scintillation detector. A 6.1 litre volume stainless steel cell was 
filled with an in-house made liquid scintillator loaded with Gd to 
enhance neutron capture. A two-pulse signature (proton recoils followed
by gammas from neutron capture) was used to identify the neutron 
events from much larger gamma background from PMTs. Suppression of 
gammas from the rock was achieved by surrounding the detector with 
high-purity lead and copper. Calibrations of the detector were 
performed with various gamma and neutron sources. Special care was 
taken to eliminate PMT afterpulses and correlated background events 
from the delayed coincidences of two pulses in the Bi-Po decay chain.
A four month run revealed a neutron-induced event rate of 
$1.84 \pm0.65 (stat.)$ 
events/day. Monte Carlo simulations based on the GEANT4 toolkit 
were carried out to estimate the 
efficiency of the detector and the energy spectra of the expected proton
recoils. From comparison of the measured rate with Monte Carlo simulations
the flux of fast neutrons from rock was 
estimated as 
$(1.72 \pm 0.61 (stat.) \pm 0.38 (syst.))\times 10^{-6}$~cm$^{-2}$~s$^{-1}$
above 0.5~MeV.

\end{abstract}

\end{center}

\vspace{0.5cm}


\vspace{0.5cm}
\noindent Corresponding author: E. Tziaferi,
Department of Physics and Astronomy, University of Sheffield, 
Hicks Building, Hounsfield Road, 
Sheffield S3 7RH, UK

\noindent Tel: +44 (0)114 2223547; \hspace{2cm} Fax: +44 (0)114 2223555; 

\noindent E-mail: e.tziaferi@sheffield.ac.uk

\pagebreak

\section {Introduction}

Neutrons are the most important background for a large variety 
of underground experiments. Experiments searching for Weakly 
Interacting Massive Particles (WIMPs), for example, work with 
a very low energy threshold, and are sensitive to (and should be 
protected from) neutrons from all possible sources: rock, detector, 
shielding components and cosmic-ray muons.
Neutrons, like WIMPs, induce nuclear recoils, so it is crucial that the
neutron flux is suppressed by shielding, active veto systems and the 
use of ultra-pure materials.

The neutron flux from rock dominates over other sources. 
It originates in spontaneous fission (mainly $^{238}$U) and 
($\alpha$,n) reactions due to uranium and thorium traces.
The suppression of this flux can be achieved by installing  
hydrocarbon shielding around the detector. The required 
thickness of the shielding is determined by 
the neutron flux and the projected sensitivity of the WIMP dark matter 
detector. The neutron flux from rock and its effect on the detector 
sensitivity can be calculated using the measured contamination levels 
of U and Th. However, there may be quite a large uncertainty in such 
a calculation. Firstly, the cross-sections of ($\alpha$,n) reactions
are not known with high precision for a large number of isotopes, in 
particular the energy spectra of emitted neutrons are quite uncertain.
Secondly, Monte Carlo codes for neutron transport need to be 
extensively tested and should use the most accurate libraries for 
neutron interaction cross-sections. Thirdly, measurements of U/Th 
concentrations may not provide accurate results for the actual 
concentrations present. If such measurements are done using 
mass-spectrometry, then only a few samples can be tested and further 
studies then rely on the assumption of a uniform concentration of 
U/Th in the rock. If they are performed with a Ge detector, then the gamma lines 
provide an estimate of concentration of certain isotopes in the U/Th 
decay chains and the evaluation of the neutron flux depends then on 
the assumption that there is equilibrium in these chains. Obviously an 
accurate calibration of the Ge detector is also necessary. Finally, the 
exact composition of the rock may not be known to a high degree of 
accuracy. For example, the fraction of water (hydrogen) may vary and 
is anyway difficult to measure. Meanwhile, hydrogen (even 1\% of H) affects 
dramatically the moderation and absorption of neutrons 
\cite{lemrani,wulandari}. So direct measurements of the neutron 
flux from rock is important as a means of checking the calculations based on 
other U/Th concentration measurements.

The Boulby Underground Laboratory situated at a vertical depth of 
1070 m underground (2800~ m~w.e. \cite{matt}) hosts
several dark matter experiments \cite{naiad, driftII, zeplinI, zeplinII}.
The concentrations of U and Th in Boulby rock which is almost pure 
NaCl salt (surrounding the underground laboratory), 
were evaluated from measurements of gamma lines 
associated with U/Th decay chains \cite{peter}. 
A Ge detector was exposed to the gamma 
flux from the rock surrounding the laboratory hall and the 
intensitities of the observed gamma lines were converted into U/Th 
concentration levels using Monte Carlo simulations \cite{peter}.

In this paper we discuss our technique to measure low-intensity neutron 
fluxes and the application of this technique to the study of the neutron 
background from rock at Boulby mine. 
The detector, scintillator and data aquisition system (DAQ)
are described in Section 2. Detector calibrations are presented in 
Section 3. Our measurements of neutron flux from the rock are described 
in Section 4. The results of the experiment are shown and
the uncertainties are discussed in Section 5. Finally the conclusions 
are given in Section 6.

\section {Detector and DAQ}
\label{detector}

The liquid scintillation detector used in the experiment consisted 
of a cylindrical 
stainless steel vessel 19.7 cm diameter and 20.0 cm length filled 
with a liquid scintillator based on
diisopropylnaphthalene (C$_{16}$H$_{20}$) 
of volumic density 0.97 g/cm$^3$, 
loaded with $\sim$0.2\% of Gd to increase the probability of neutron capture. 
The recipe used for Gd loading is similar 
to that described in Ref. \cite{phil} where much higher loading of Gd 
had been achieved. The attenuation length of the scintillator was measured to 
be about 5 metres, well exceeding the size of the detector.
The scintillator has a high flash-point (more than 100$^{\circ}$C) and 
low-toxicity - requirements imposed by health and safety considerations
of a working mine. 

\begin{figure} [htb]
\begin{center}
\includegraphics{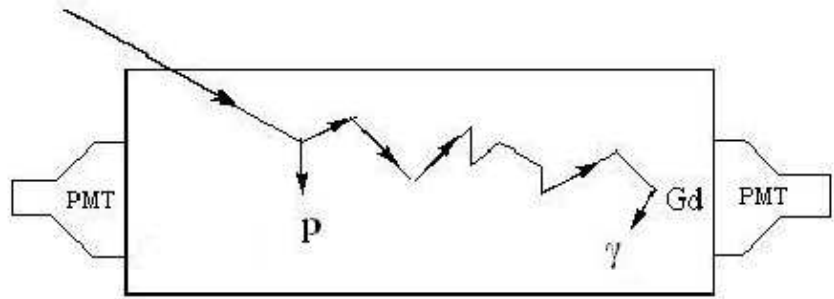}
\caption{Detection principle:  A neutron entering the detector produces 
proton recoils by elastic scattering, then slows down, is thermalised 
and finally captured with the emission of gammas.}
\label{det_priciple}
\end{center}
\end{figure}

The two flat surfaces of the detector were viewed by two 5'' 
fast Hamamatsu photomultiplier tubes (PMTs) 
through quartz windows, attached with optical coupling 
grease (Figure~\ref{det_priciple}). The inner surface of the vessel, 
including the two rings around the
quartz windows, were coated with aluminised mylar with a coefficient of
reflection greater than 0.9 to maximise light collection.

The detector was running in an environment with a temperature 
of more than 30$^{o}$C, so a cooling system was installed to prevent 
degradation of the scintillator. Copper coils were wrapped around the
vessel and running water, from a chiller, enabled the temperature 
to be maintained at about 20$^{o}$C. 

The detector was installed inside a `castle' made of ultra-pure copper 
and lead and N$_2$ was bubbled through the scintillator 
to remove dissolved oxygen. The temperatures of the scintillator
(about 20$^{o}$C), the castle  (26$^{o}$C) and the ambient air
(30-33$^{o}$C) were measured with thermocouples. 
The thickness of lead and copper in the `castle' was 7.5 and 10 cm, 
respectively, for the top and three vertical sides of the castle. For the 
bottom part and one vertical side the thickness of lead was 15 cm. 
Lead and copper shielding ensures the suppression of the gamma flux 
from the rock by several orders of magnitude. For energy calibrations 
with gamma sources a hole was made in the roof of the castle into 
which a source could be inserted. With the hole plugged none of the 
gamma sources could be detected.
The configuration of the detector with shielding and the rest
of the environment was included in our Monte Carlo code to simulate the 
detector response to various radiations.

\begin{figure} [htb]
\begin{center}
\includegraphics{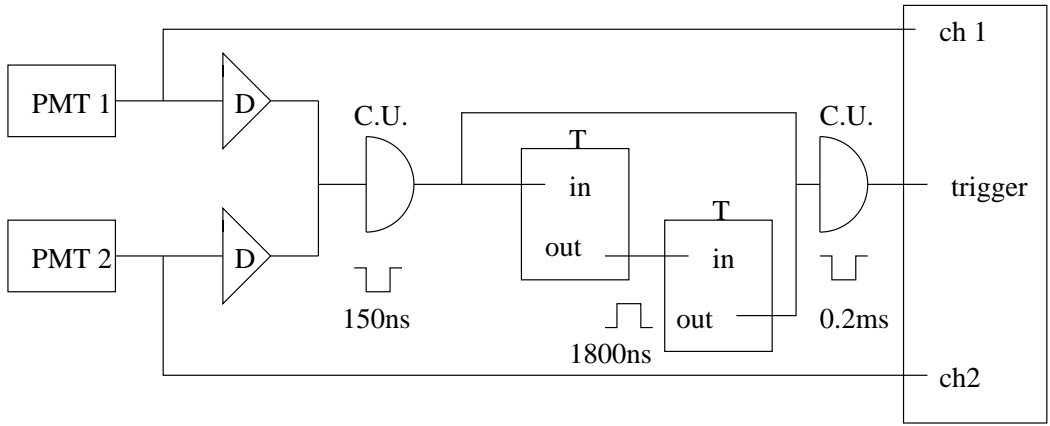}
\caption{The functional diagram of the electronics for neutron measurement and neutron calibration (2 pulses in delayed coincidences). The discriminators are labeled as 'D', the coincidence units as 'C.U.' and the timers as 'T'.}
\label{daq}
\end{center}
\end{figure}

Two different DAQ circuits were used: one for gamma calibration 
(giving 1 pulse) and one  
for neutron measurement (giving 2 pulses in delayed coincidences),
see Figure~\ref{daq}. 
For all runs, the signals from both PMTs were fed into NIM low-threshold 
discriminator units set to trigger when the signal exceeded a threshold
of 10~mV, equivalent to 2 photoelectrons from each tube. 
For energy calibration runs, the logic signals from the
discriminator channels were fed into a coincidence unit which 
triggered the waveform digitisers (CompactPCI Acqiris digitiser) when 
signals arrived from the two PMTs (time gate 150 ns). When the
waveform digitiser was triggered, the signals from both PMTs were
digitised (sampling rate 0.5 GHz, digitisation accuracy 2 ns, 
total digitisation time 200 ns), 
passed to a computer and stored on disk for off-line analysis.

For neutron measurements and calibrations, the output signal of the 
coincidence unit was split in two, one fed into the second coincidence 
unit and the other sent to two timers sequentially. The first timer 
gave a 1.8 $\mu$s delay and the second timer opened a time gate of
200 $\mu$s. The gate from the second timer was sent to the second 
coincidence unit and this unit generated a trigger if the second 
pulse arrived within the 200 $\mu$s time gate. 
The trigger was sent to the digitiser and the
signals from the PMTs were digitised in the same way as described above 
but the total digitisation time was 200 $\mu$s. Such a scheme 
thus enabled detection of two pulses - the first one assumed to be 
from proton recoils induced by neutron elastic scattering and the 
second one due to the Compton electrons produced by gammas from 
neutron capture on Gd, H or other elements. The first timer with 1.8 $\mu$s 
delay was needed to avoid triggering on the 1st pulse twice (the 2nd 
pulse should be delayed relative to the 1st one by a certain time).

\section {Calibrations}\label{calibrations}

To characterise the detector and its response to various radiations 
the following calibrations were carried out: 
\begin{enumerate}
    \item Energy calibration using gamma-rays from $^{57}$Co, 
    $^{137}$Cs and $^{60}$Co radioactive sources. 
    \item Calibration with neutrons from a $^{252}$Cf neutron source 
    with the detection of delayed coincidences 
    to demonstrate the sensitivity of the detector to neutrons 
    and to check efficiency.
    \item Calibration with gamma-rays from $^{60}$Co source with the 
    detection of delayed coincidences to show that these coincidences 
    produce a random background -- not connected to neutrons or any 
    spurious effects.
\end{enumerate}

In the absence of photoelectric peaks from gamma-ray lines in 
light materials, the energy calibration of the detector was performed 
using the Compton edges of the energy spectra of events from 
different sources. Due to the finite energy resolution of the 
detector the Compton edge was smeared out and accurate calibration 
was possible only by comparing the measured energy distribution with 
simulations. Simulations were carried out using the GEANT4 
toolkit \cite{geant4} taking into account the configuration of the 
detector, shielding and the position of the source. The simulated 
spectrum of energy depositions was convolved with the energy 
resolution of the detector assumed to be Gaussian with standard 
deviation of the form $(\sigma / E)^2 = \alpha + \beta / E$ 
\cite{birks}. The parameters $\alpha$ and $\beta$, as 
well as the conversion factor from the measured charge to the 
energy scale, were determined from a comparison with the measured 
spectrum. 

Three gamma-ray sources were used for energy calibration:
$^{57}$Co (most intense gamma-ray line at 122~keV, 
40~keV Compton edge), $^{137}$Cs (622 keV gamma-ray line, 480~keV 
Compton edge) and $^{60}$Co (1173~keV and 1332~keV gamma-ray lines, 
about 1040~keV mean energy for the Compton edges). The $^{57}$Co 
source has a Compton edge very close to the energy threshold of 
the detector. In this case the Compton edge could be confused with (and/or 
superimposed on) the threshold effect. To avoid errors, the energy 
spectrum from this source was used only to check 
the energy calibration with other sources. The calibrations with all 
sources showed consistent results with the values of $\alpha$ = 0.08 and 
$\beta$ = 3.9 keV if $E$ is measured in keV.

\begin{figure} [htb]
\begin{center}
\includegraphics[scale=0.8]{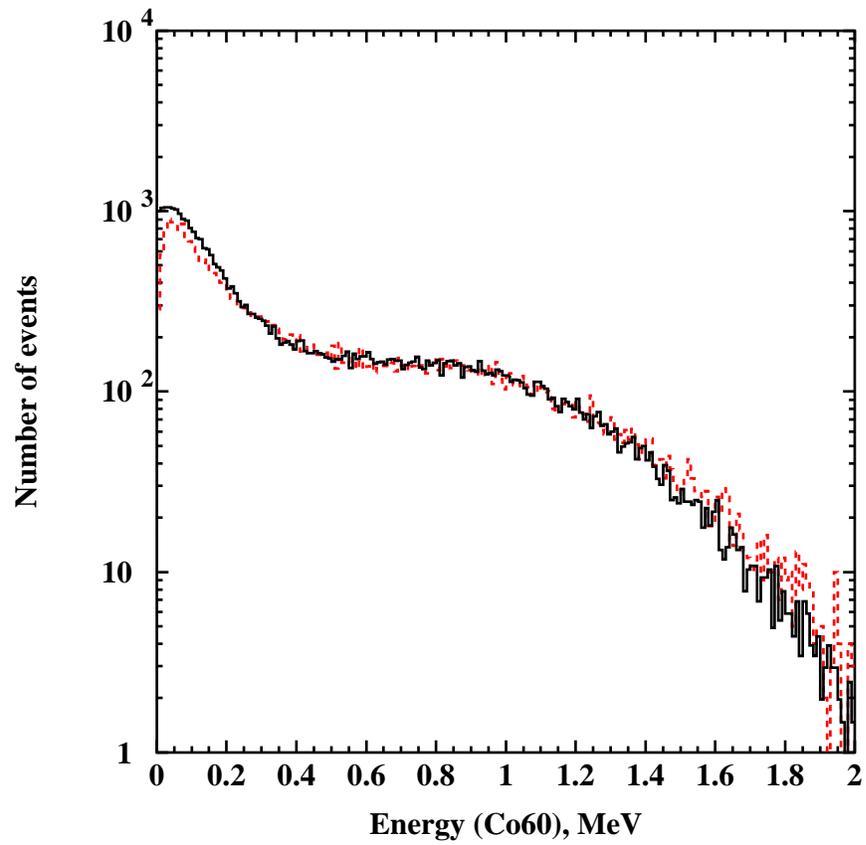}
\caption{Energy spectrum of events from $^{60}$Co: data (dashed histogram) 
and GEANT4 simulations (solid histogram).}
\label{encal}
\end{center}
\end{figure}

Energy calibration of the detector with different sources was 
carried out regularly with time intervals of 1 - 2 months and the 
conversion factor found to be stable within 10\% over the 4 month 
running time of the experiment. An average value was used in the analysis. 
Possible uncertainties in the conversion factor and energy resolution 
and their effects on the neutron flux measurements are 
discussed in Section~\ref{uncertainties}. An example of the spectrum 
of energy depositions from $^{60}$Co is shown in Figure~\ref{encal}.

\begin{figure} [htb]
\begin{center}
\includegraphics[scale=0.8]{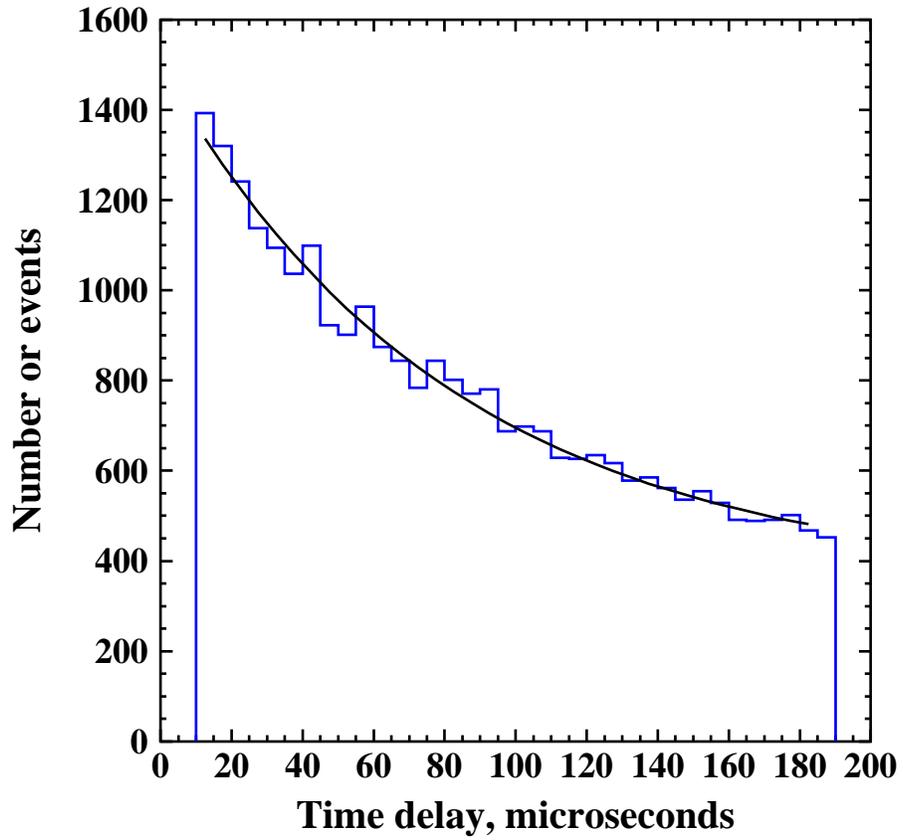}
\caption{Distribution of time delay between the 1st (proton recoils) 
and 2nd (neutron capture) pulses in events from $^{252}$Cf data. 
A fit with a sum of an exponential and a constant is also shown with
a solid curve.}
\label{timedelay}
\end{center}
\end{figure}

Neutrons from spontaneous fission of $^{252}$Cf were used to test 
if the detector was sensitive to neutrons and to estimate
the efficiency of the detector. The source produced a total rate of 
about 13000 neutrons per second and was 
positioned either at a distance of about 1 m from the detector or 
on top of the 'castle'. As described in Section~\ref{detector} the DAQ 
was configured to detect two pulses in delayed 
coincidence within a time window of 200~$\mu$s 
-- the first pulse due to proton recoils from 
neutron elastic scattering and the second due to gammas 
from neutron capture. As the incident neutron flux from the $^{252}$Cf 
source is very high, there was a non-zero probability of detecting 
random coincidences within 200~$\mu$s between two pulses produced by 
proton recoils from two neutrons or two pulses produced by gammas 
from two captured neutrons. Figure \ref{timedelay} shows the time 
delay distribution of the two pulses. In the calibration the 
presence of random coincidences described above adds a flat 
distribution superimposed on the exponential. (Strictly speaking 
the component due to random coincidence should also be exponential,
but since the mean time between two random pulses always exceeded
4~ms, this exponential can be approximated by a constant on a time scale
of 200~$\mu$s.) Each distribution was fitted with the two-component function 
$N = N_{n} /\tau \times \Delta t \times \exp(-t/\tau) + N_{b}$ 
with 3 free parameters: the number of background events per time bin 
$N_{b}$, the mean time delay (index of the exponential) $\tau$ and the 
total number of neutrons $N_{n}$. In the formula above, $\Delta t$ 
denotes the width of the time bin and $N$ denotes the number of 
events per bin. Note that $N_{n}$ is the total number of neutrons 
integrated over all time delays, so a correction for reduced 
efficiency due to reduced range of time delays is not needed. 
The index of the exponential  was found using a fit to the 
neutron calibration data which had the highest statistics: 
$\tau = 84.21 \pm 5.50$~$\mu$s. 

The rate of delayed coincidences between proton 
recoils and gammas from neutron capture was found to be $5.87 \pm 
0.68$~s$^{-1}$  and $0.53 \pm 0.04$~s$^{-1}$ during the calibration run 
having the source on top of the 'castle' and 1 m away respectively 
(for 50 keV energy threshold
for the 1st pulse and 100 keV threshold for the 2nd pulse -- 
the reasons for these
thresholds will be given below).

\begin{figure} [h!]
\begin{center}
\includegraphics[scale=0.8]{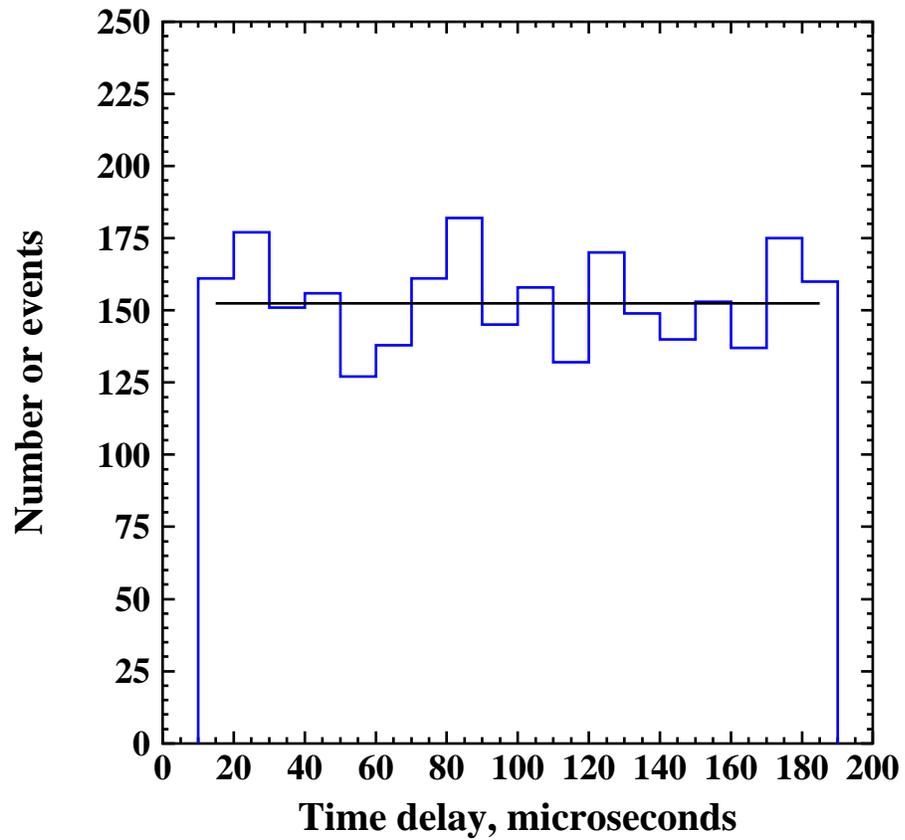}
\caption{Distribution of time delay between the 1st and 2nd pulses 
in events from $^{60}$Co source due to random coincidences between gammas. 
A fit with a constant function is also shown with a solid curve.}
\label{tdelayco}
\end{center}
\end{figure}

In addition, a calibration run with the $^{60}$Co source was carried out with 
the same DAQ settings as for the run with $^{252}$Cf. The 
time delay distribution  is shown in Figure \ref{tdelayco}. 
The presence of the two pulses in an event in this case is due to 
random concidences of two gammas within a time window of $200$~$\mu$s. 
The time delay distribution is flat proving that the exponential shape 
observed with the neutron source is not an effect of DAQ but is due 
to neutron events. The results from these calibrations proved 
that the detector was sensitive to neutrons and allowed the  
efficiency of neutron detection to be calculated.

\begin{figure} [htb]
\begin{center}
\includegraphics[scale=0.8]{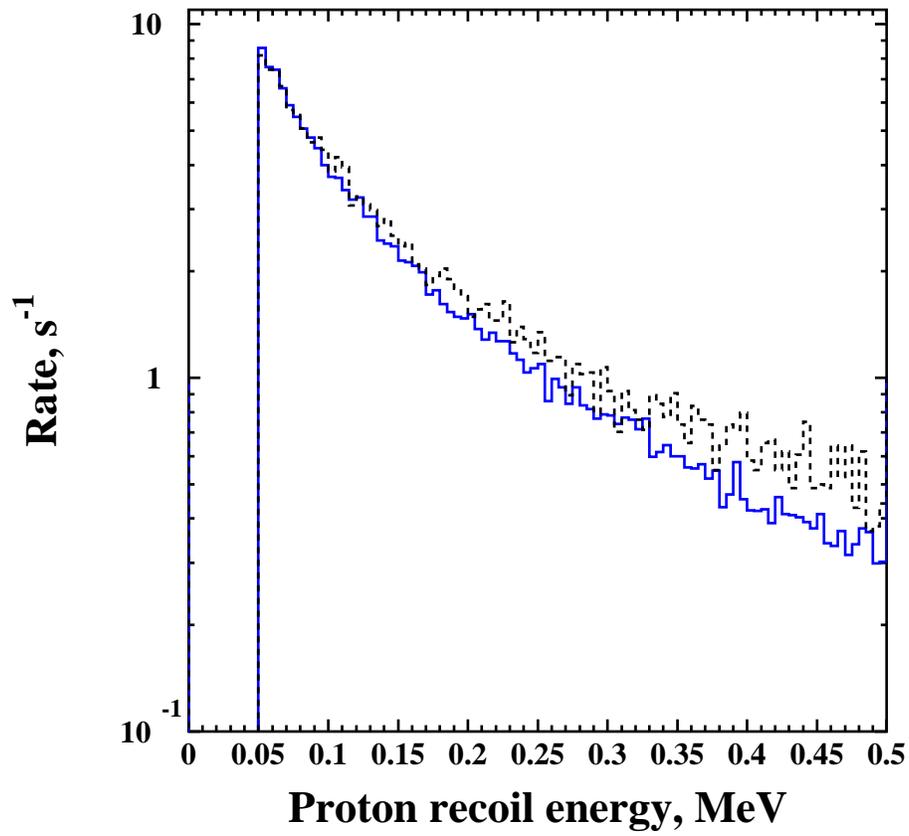}
\caption{Proton recoil spectra (rates in absolute units per 5~keV
without any normalisation) from data (dashed histogram) and 
GEANT4 simulations (solid histogram) of $^{252}$Cf run.}
\label{rate_protonrec}
\end{center}
\end{figure}

Finally, a $^{252}$Cf run was carried out having the same DAQ 
settings as for the energy calibration runs, recording single pulses 
(mainly proton recoils with some admixture of gammas), 
without delayed coincidences. The energy spectrum of proton 
recoils  as measured with the neutron source on top 
of the castle is shown in Figure~\ref{rate_protonrec} (dashed line). 
Also the corresponding spectrum (without any normalisation) 
calculated with GEANT4 (solid line) is shown, taking into account 
typical quenching of scintillations for protons in organic liquid 
scintillators \cite{quenching} in the form: $E = 0.2 
E_{p}^{1.53}$, where $E_{p}$ is the kinetic energy of a proton in MeV
and $E$ is the measured energy in MeV relative to the high-energy gamma
calibration point from the $^{60}$Co source.
The small difference, at high energies between the measured and 
simulated spectra is likely due to the presence of gammas from neutron 
capture in the calibration data (see the explanation above).
In the simulation no neutron captures were included. 
The measured rate was $163.88 \pm 1.40$~s$^{-1}$ and the simulated one 
was 140.30~s$^{-1}$, for the energy range of 50-500~keV. 
This agreement allows us to conclude that GEANT4 simulates neutron
transport and production of proton recoils with reasonable accuracy. 
The uncertainty can be estimated from the difference between measured
and simulated rates which is 14\%. This is consistent with the 
conclusion of Ref.~\cite{lemrani} that claims a good agreement between 
GEANT4~\cite{geant4}, MCNPX~\cite{mcnpx1, mcnpx2} and GEANT3~\cite{geant3} 
for neutron propagation through large thicknesses of rock and shielding.

The calibration runs described above allowed determination of the 
conversion factor between the detected charge and the energy of 
particles (electron equivalent energy), the energy resolution of the 
detector, the energy threshold and the trigger efficiency as a 
function of measured energy (the later is discussed in 
Section~\ref{uncertainties}). The results of these runs proved that 
the detector is sensitive to neutrons and that the Monte Carlo 
simulations using the GEANT4 toolkit were able to 
reproduce the energy spectrum and absolute rate of proton recoils.
Comparison between the measured rate of delayed coincidences and the 
measured/calculated rates of single proton recoils allowed us to estimate
the efficiency of detecting delayed coincidences (see below).

\section {Measurements of the neutron flux from rock} \label{measurements}

The experiment to measure the neutron flux at the Boulby Undergound 
Laboratory was carried out from November 2004 to September 2005. 
The first three months of the experiment were dedicated to studying 
the behaviour of the detector and its  background. Measurement 
of the neutron flux from the rock were  performed between the end of 
February 2005 and the end of June 2005. 
During that time all DAQ settings remained the same and the run 
was interrupted only to carry out various calibrations of 
the detector to ensure its stability.

The neutron flux from salt rock was expected to be quite low requiring 
sensitive equipment for its measurement. An important part of the 
experiment was the study and rejection of background events which 
might be an obstacle for such a sensitive measurement. 

Two important sources of background events producing delayed 
coincidences were found in this experiment. The first are 
afterpulses from the PMTs, known to occur up to a 
few microseconds after the main pulse from the PMT. Their amplitude 
normally does not exceed a few photoelectrons, which is equivalent 
to a few tens of keV for our detector. The probability of having two 
afterpulses from the two PMTs during the 150~ns coincidence gate 
used here is very low, however the experiment 
in fact does have a low signal rate, the total rate being 
a few tens of events per day. The rate of random coincidences between 
afterpulses and/or noise pulses may not be negligible compared to the 
signal rate. A histogram of time differences between two 
secondary pulses from the two PMTs for such events 
is flat showing that secondary pulses from the two PMTs are not 
correlated but are due to random coincidences between afterpulses
and/or noise pulses from different PMTs. 
Thus, such a background is due to the delayed coincidence of the following 
pulses: first pulse - normal scintillation pulse on both PMTs; second
pulse - afterpulses and/or noise pulses on two PMTs.
The rejection of this background was done by 
requiring the second pulse to have an amplitude of more than 100 keV 
and to be delayed by more than 10~$\mu$s relative to the first pulse. 
The fraction of events from $^{252}$Cf which passed the above cut 
is (75 $\pm$ 1)\%.
Such a cut limits the sensitivity of the detector to neutrons.
The same cut was applied to data and calibration run with $^{252}$Cf
and its efficiency was automatically included in the overall efficiency
of detecting the second pulse in an event. This efficiency does not 
depend on the source of neutrons, either Cf or rock, because this
cut removes only second pulses induced by gammas from neutron captures,
whose spectrum does not depend on the neutron source. 

\begin{figure} [htb]
\begin{center}
\includegraphics[scale=0.8]{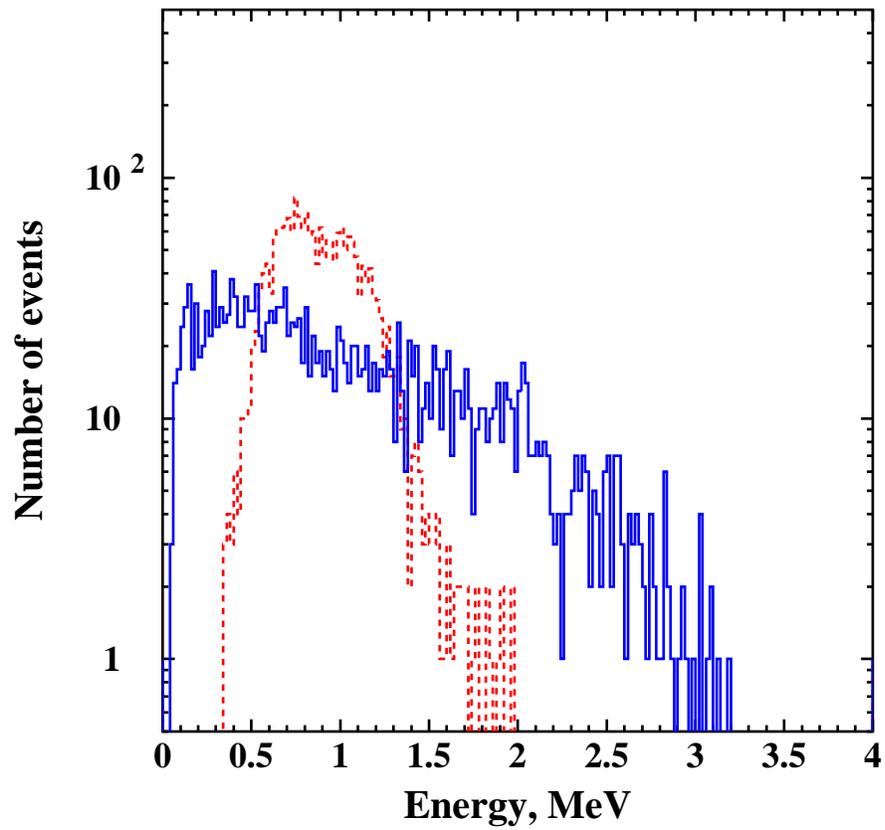}
\caption{Measured energy spectra of the first pulse (solid histogram) from 
beta emission and the second pulse (dashed histogram) from alpha decay, 
for the correlated background due to the Bi$\rightarrow$Po$\rightarrow$Pb 
decay.}
\label{corrsp}
\end{center}
\end{figure}

\begin{figure} [htb]
\begin{center}
\includegraphics[scale=0.8]{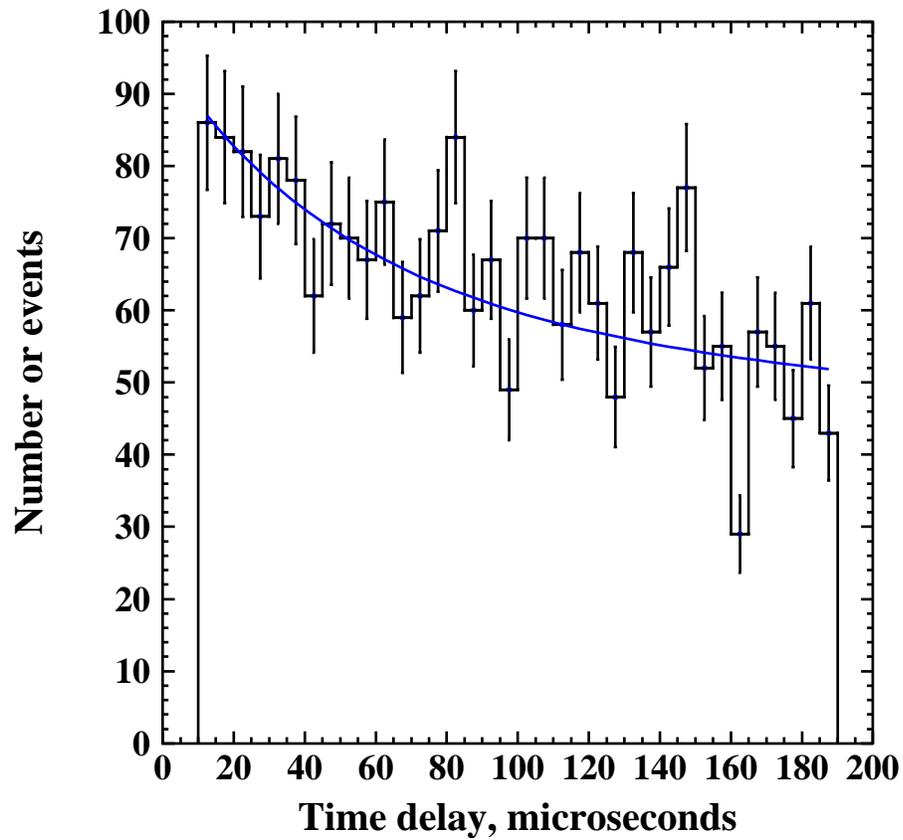}
\caption{Measured distribution of the time delay between the first (from beta 
emission) and second (from alpha decay) pulses in the events 
from the correlated background
due to the Bi$\rightarrow$Po$\rightarrow$Pb decay.
The solid curve is the fit with an exponential of 237~$\mu$s.}
\label{corrtdelay}
\end{center}
\end{figure}

The second source of background arises from the possibility of 
two consecutive decays 
in the $^{238}$U decay chain: $^{214}$Bi$\rightarrow ^{214}$Po
$\rightarrow ^{210}$Pb. The first decay goes via beta emission 
whereas the second one is an alpha-decay. The half-life of $^{214}$Po 
is 164~$\mu$s which makes the two pulses from this decay chain 
capable of mimicking the signal from neutrons. Prior to 
installation of the detector, the scintillator was exposed to air 
with a certain concentration of radon. Radon daughters were diffused 
in the scintillator and at the beginning of the running time (November 
-- December 2004) the rate of correlated background events from 
Bi$\rightarrow$Po$\rightarrow$Pb decays was quite high 
(the rate on the first day was 324 $\pm$ 18~events/day).
Figure 
\ref{corrsp} shows the energy spectra of the first and second 
pulses in the events from this correlated background. The energy 
spectrum of the second pulse has a clear peak at about 0.8 MeV -- 
quenched peak of alphas at 7.68 MeV. Note that the expected alpha 
peak, taking into account the quenching of alphas in standard liquid 
scintillator, should be at about 0.96 MeV~\cite{verbinski}, in a 
reasonable agreement 
with the measured value. The difference between the measured value 
for the peak position and the predictions can be used to estimate 
the uncertainty in the energy scale for proton recoils (see Section 
\ref{uncertainties}). Figure \ref{corrtdelay} shows the
distribution of the time delay between the first and second 
pulses in the events from the correlated background. It is well 
fitted with an exponential of 237~$\mu$s consistent with the origin of this 
background being due to the Bi$\rightarrow$Po$\rightarrow$Pb decays. 

\begin{figure} [htb]
\begin{center}
\includegraphics[scale=0.8]{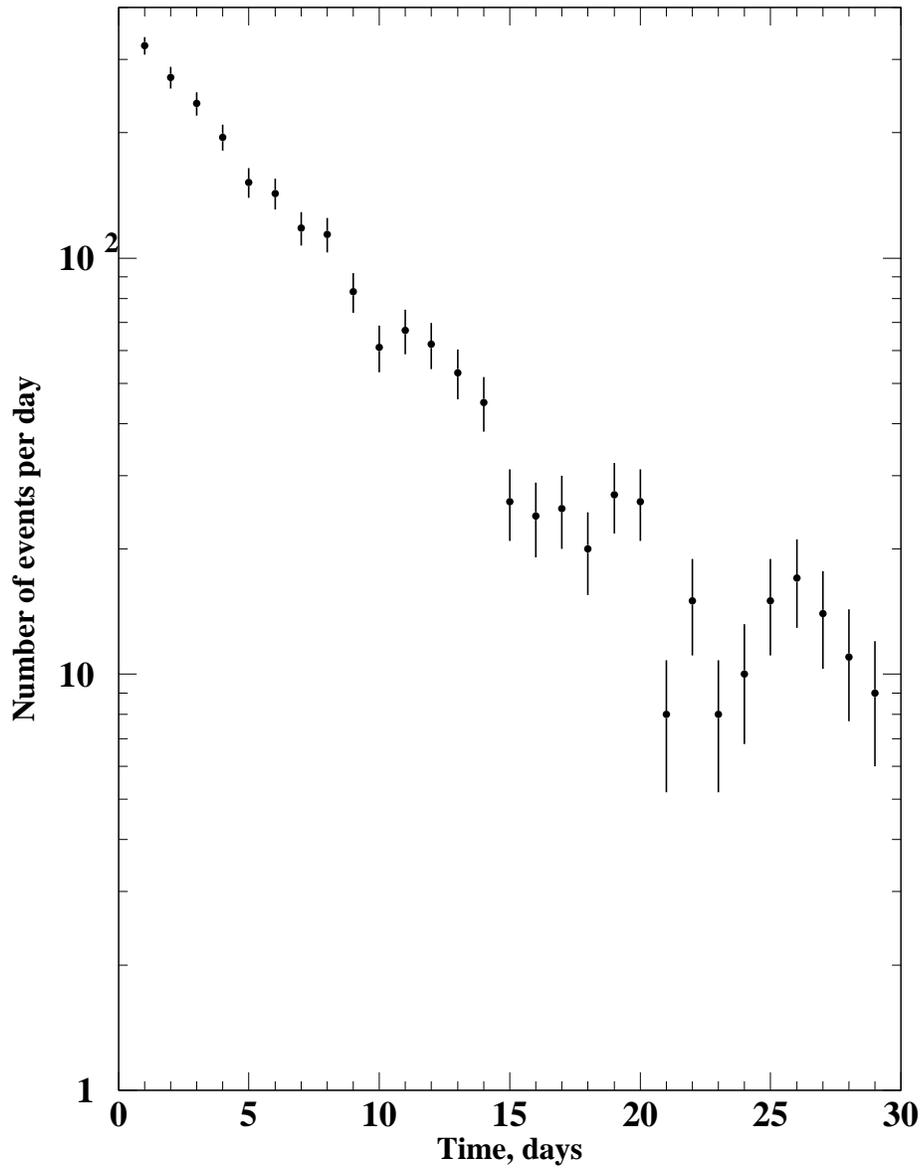}
\caption{The rate of the correlated background as a function of time
for the first month after the installation of the detector underground.
These data were not included in the analysis of neutron flux.}
\label{corrate}
\end{center}
\end{figure}

\begin{figure} [htb]
\begin{center}
$\begin{array}{c@{\hspace{5mm}}c}
\includegraphics[width=7cm]{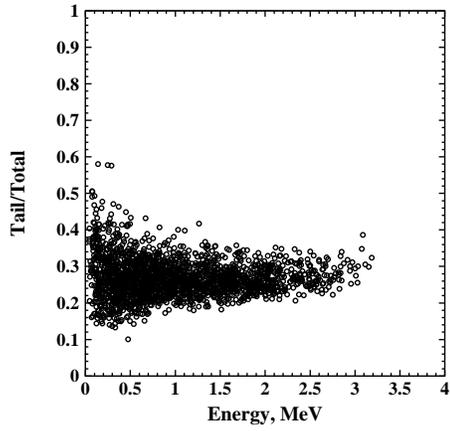} &
\includegraphics[width=7cm]{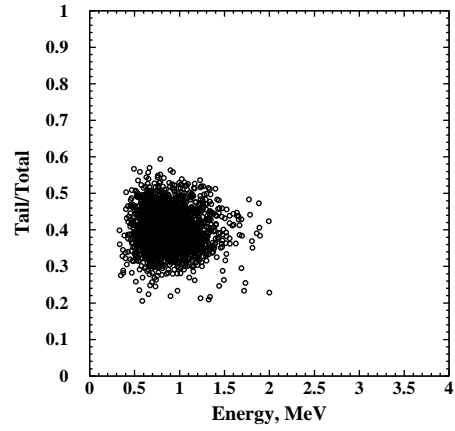} \\
\hspace{1cm} \mbox{\small{(a)}} & \mbox{\small{(b)}}
\end{array}$
\caption{The ratio of charge in the tail of the pulse to the total charge 
for the first (a) and second (b) pulses for the correlated 
background events (Bi$\rightarrow$Po$\rightarrow$Pb decay).}
\label{tailtotal}
\end{center}
\end{figure}

Figure \ref{corrate} shows the rate of correlated background 
events as a function of time since the beginning of the experiment in 
November 2004. The rate dropped to a few events per day after 1 
month of operation. The remaining events are due to intrinsic 
contamination of scintillator and/or vessel walls by uranium 
(note only isotopes in a few microns of the walls in contact with 
scintillator can produce an alpha pulse in the scintillator). 
These events can be 
eliminated using the following cuts. Firstly, the fact that the 
secondary alpha events are concentrated around 0.8 MeV can be used 
to reject this background. However, as can be seen from Figure \ref{corrsp}
the alpha peak is very broad and many neutron capture events 
also have energies in this range. Secondly, the correlated background 
events can be rejected using pulse shape analysis: the second 
pulses are due to alphas in the correlated background events whereas 
for neutron events they are due to Compton 
electrons. The possibility of using pulse shape 
discrimination can be seen from Figure \ref{tailtotal} which 
shows the ratio of charge in the tail of the pulse (20~to 100~ns 
since the beginning of the pulse) to the total charge.
A population of events due to alphas is clearly seen for tail/total 
ratio more than 0.2 (for the second pulses in the events -- 
Figure \ref{tailtotal}b). Pulses due to electrons (first pulse in 
the background events -- Figure \ref{tailtotal}a) 
lie mainly below 0.3 at energies above 0.5 MeV.
For further analysis only events with second pulses which had 
tail/total ratio of less than 0.2 in the energy region of 0.35 - 2.00~MeV
were used. Again, to estimate the efficiency of this selection criterium, the 
same cut was applied to the neutron calibration data. As only pulses 
from the neutron capture gammas are affected, the efficiency
does not depend on the neutron spectrum.

\begin{figure} [htb]
\begin{center}
\includegraphics[scale=0.8]{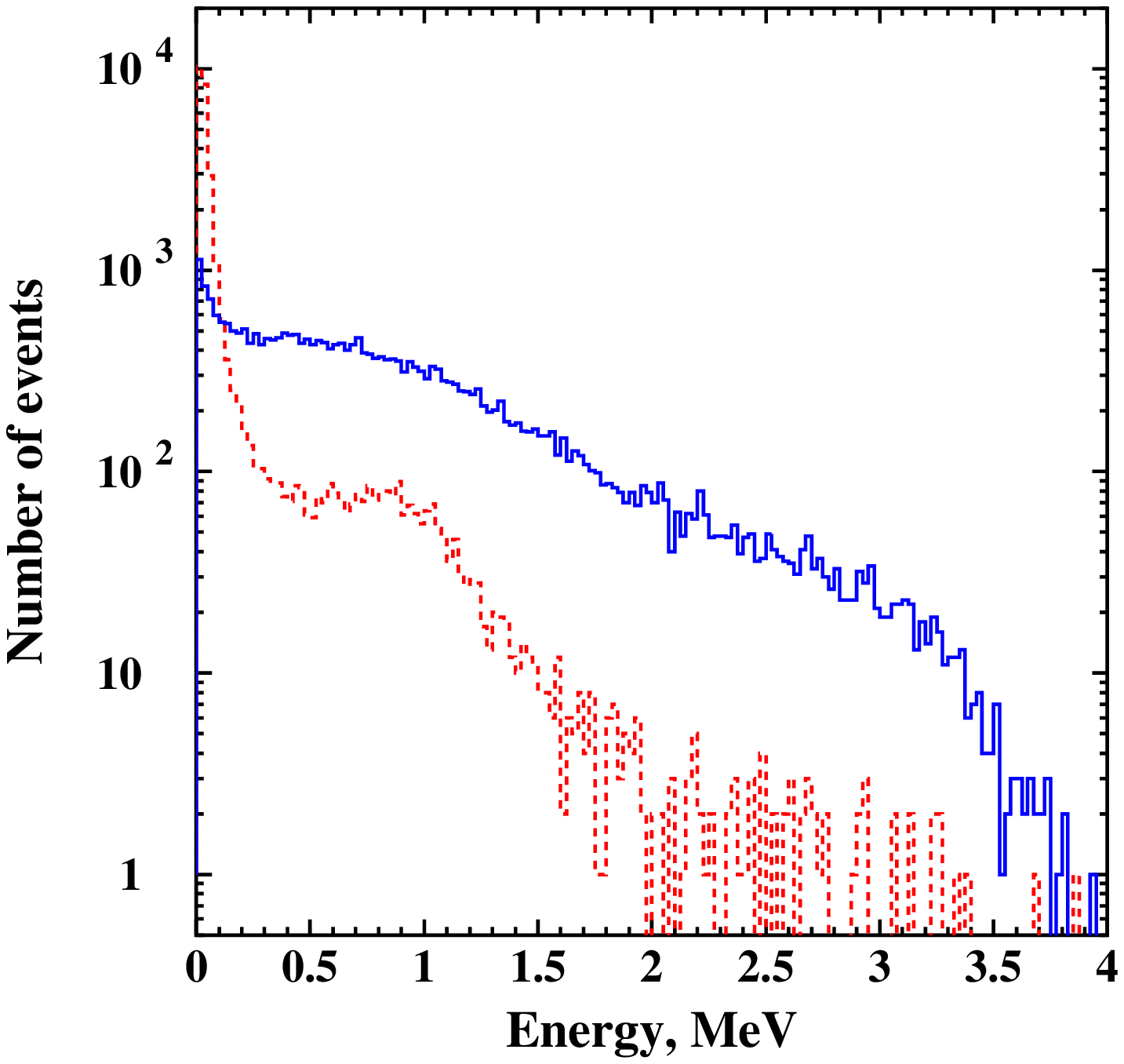}
\caption{Energy spectra of the first (solid histogram) and 
the second (dashed histogram) pulses in the events 
included in the analysis for neutron flux measurements, prior to the 
applied cuts.}
\label{datasp_nocuts}
\end{center}
\end{figure}

\begin{figure} [htb]
\begin{center}
\includegraphics[scale=0.8]{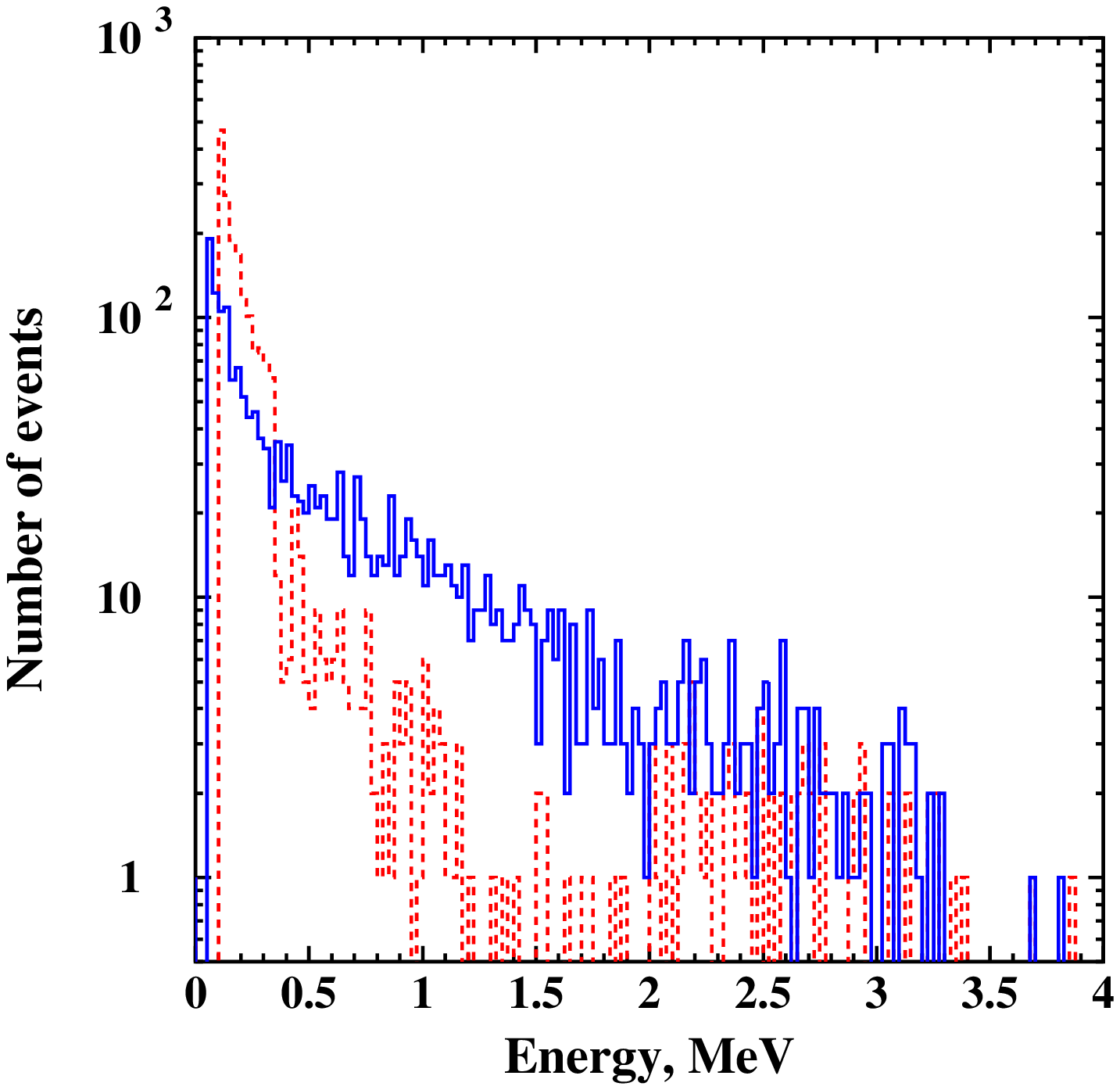}
\caption{Energy spectra of the first (solid histogram) and 
the second (dashed histogram) pulses in the events 
included in the analysis for neutron flux measurements, after the
applied cuts.}
\label{datasp}
\end{center}
\end{figure}

\begin{figure} [htb]
\begin{center}
$\begin{array}{c@{\hspace{5mm}}c}
\includegraphics[width=7cm]{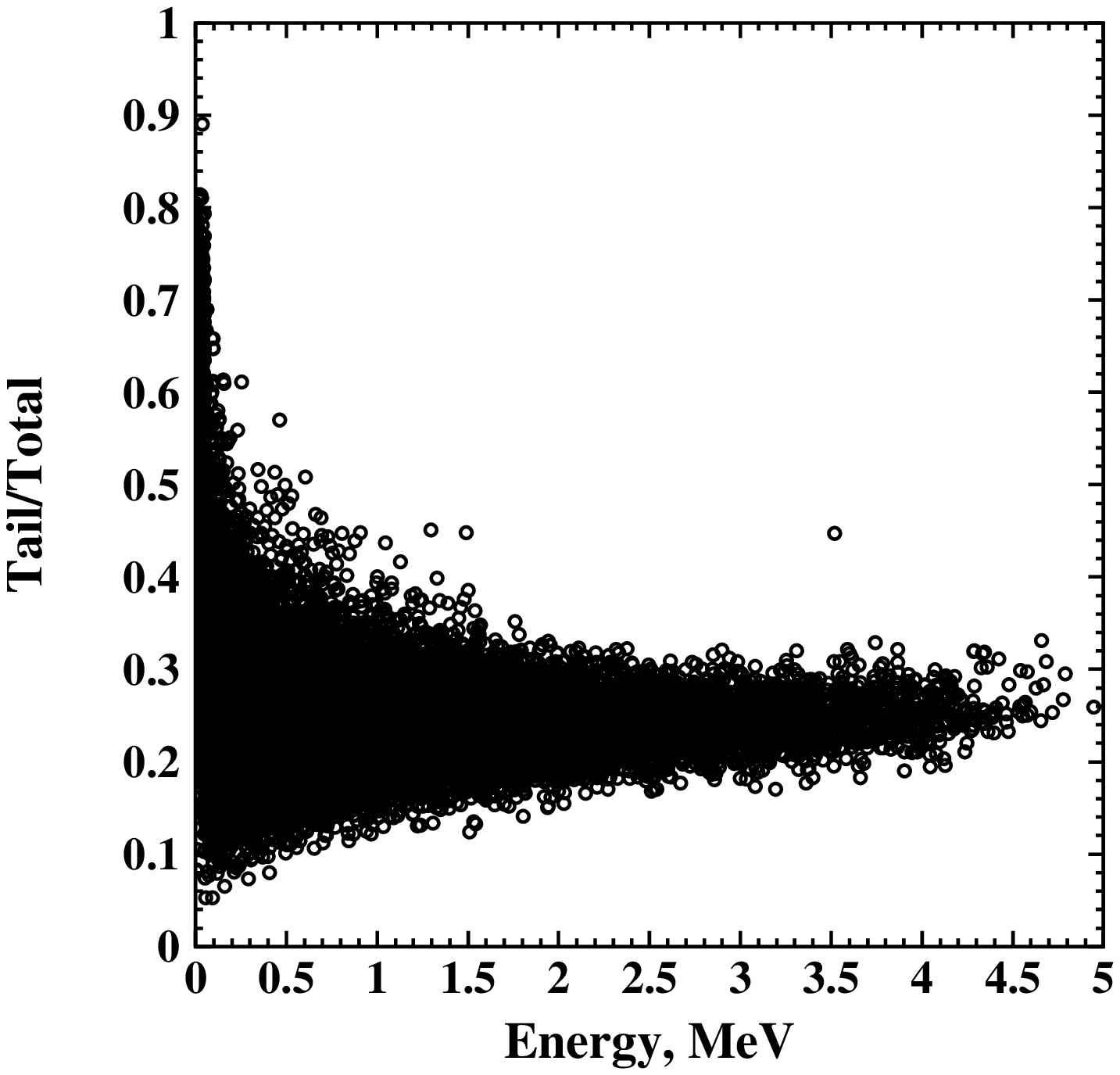} &
\includegraphics[width=7cm]{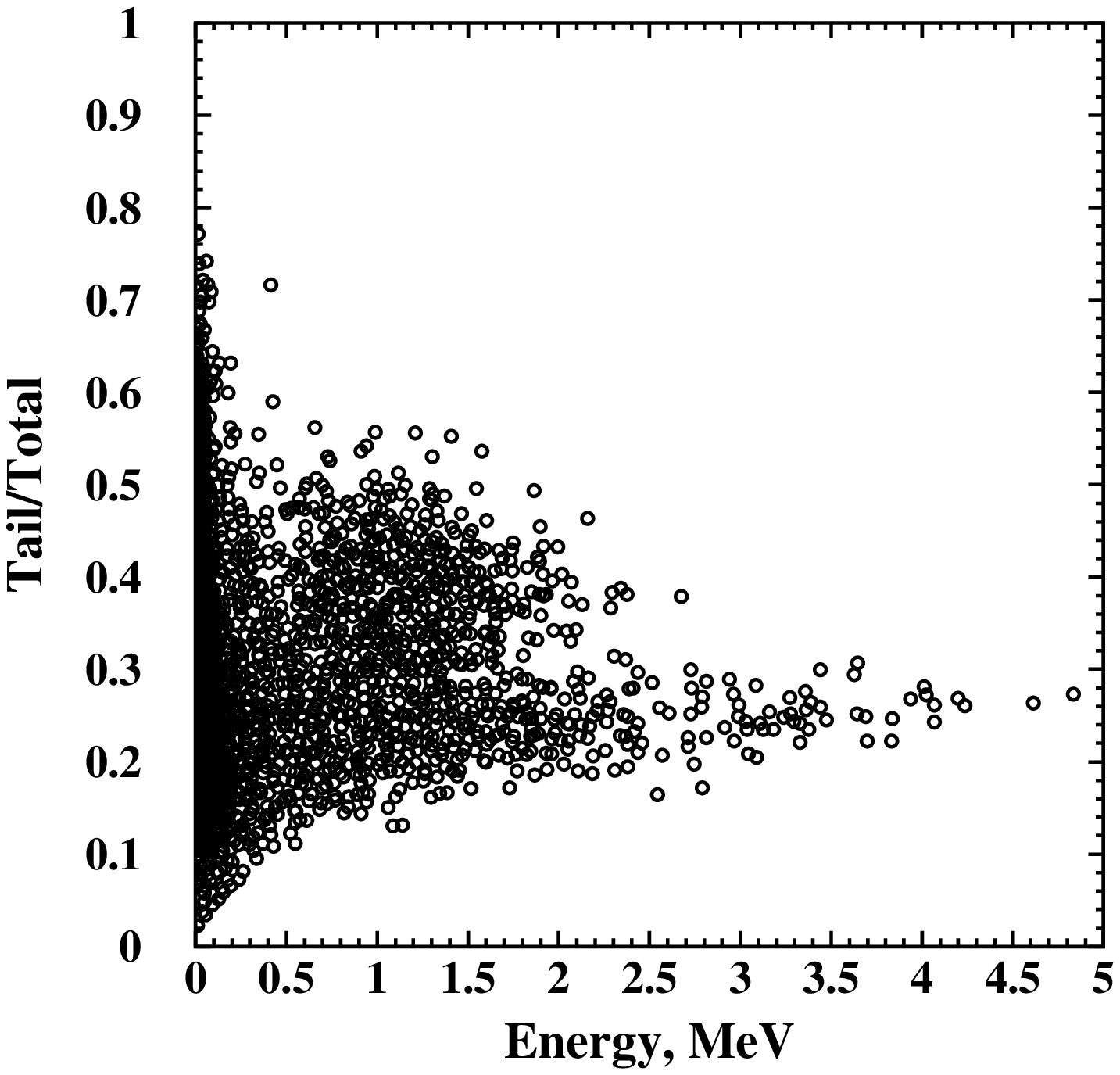} \\
\hspace{1cm} \mbox{\small{(a)}} & \mbox{\small{(b)}}
\end{array}$
\caption{The ratio of charge in the tail of the pulse to the 
total charge for the first (a) and the second (b) pulses 
in the events included in the analysis of neutron flux, prior to the 
applied cuts.}
\label{datatail_nocuts}
\end{center}
\end{figure}

\begin{figure} [htb]
\begin{center}
$\begin{array}{c@{\hspace{5mm}}c}
\includegraphics[width=7cm]{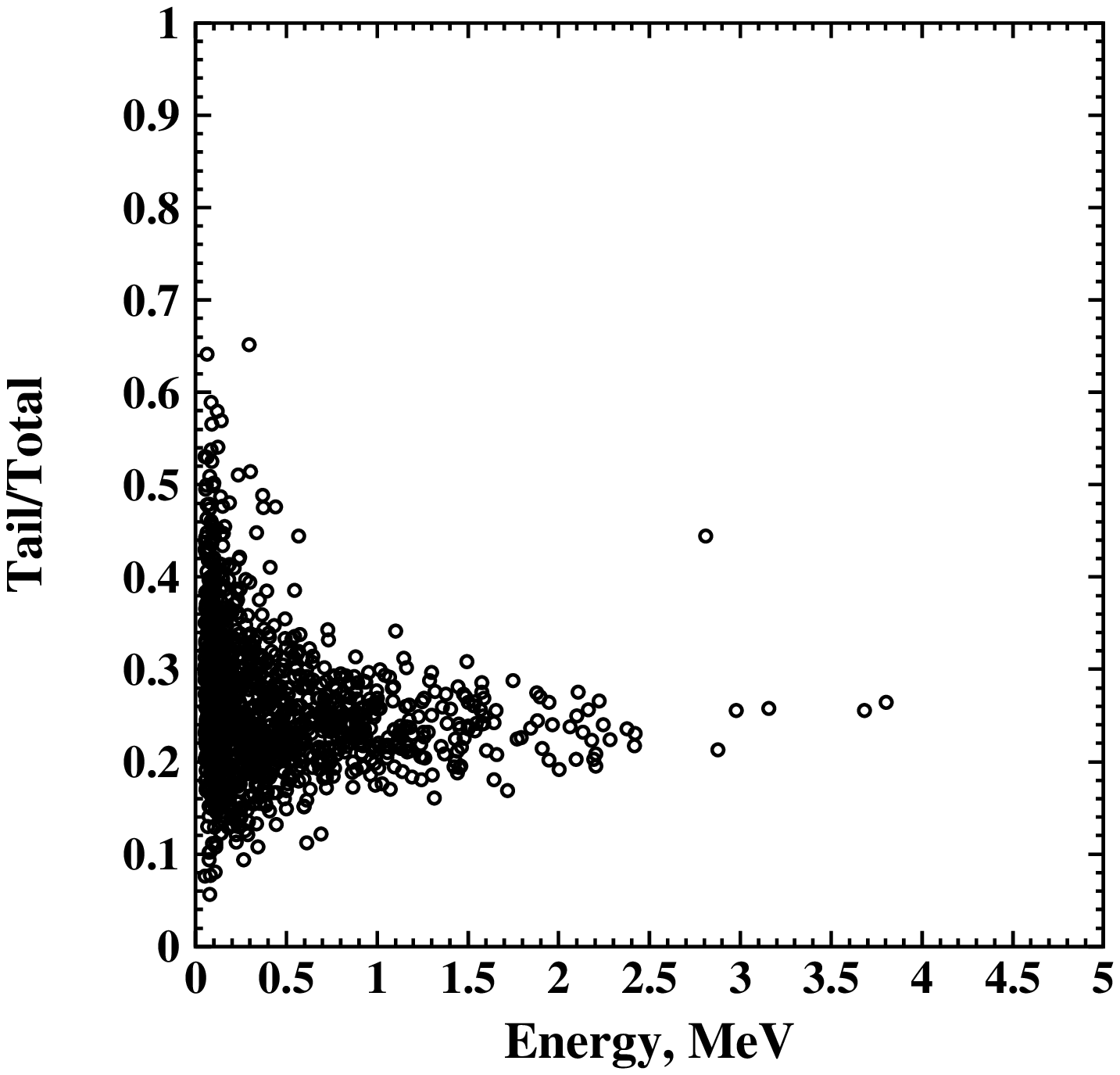} &
\includegraphics[width=7cm]{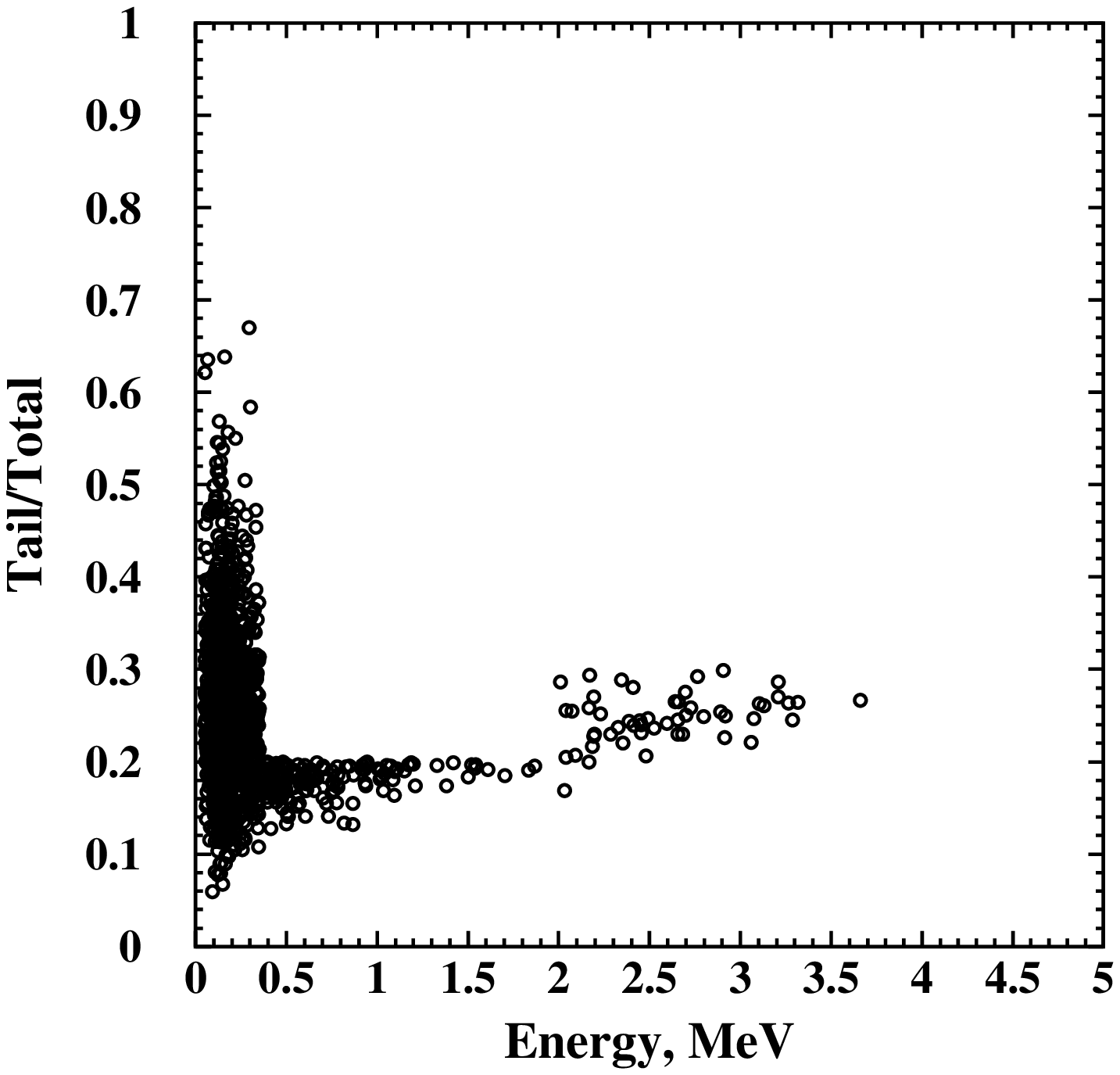} \\
\hspace{1cm} \mbox{\small{(a)}} & \mbox{\small{(b)}}
\end{array}$
\caption{The ratio of charge in the tail of the pulse to the 
total charge for the first (a) and the second (b) pulses 
in the events included in the analysis of neutron flux, after the
applied cuts.}
\label{datatail}
\end{center}
\end{figure}

\begin{figure} [htb]
\begin{center}
$\begin{array}{c@{\hspace{5mm}}c}
\includegraphics[width=7cm]{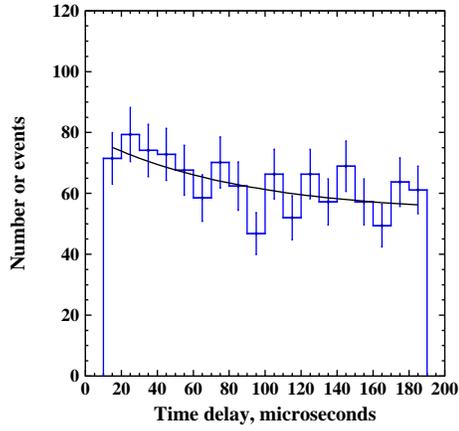} &
\includegraphics[width=7cm]{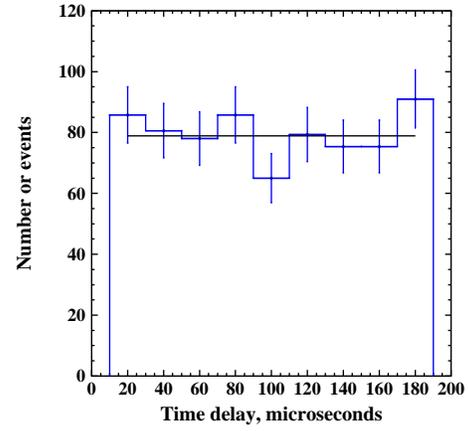} \\
\hspace{1cm} \mbox{\small{(a)}} & \mbox{\small{(b)}}
\end{array}$
\caption{Time delay distributions for the run without neutron 
shielding (a) and for the run with polypropylene shielding (b). 
The solid curve in (a) is the fit with a two-component function, 
described in the text, with fixed mean time delay. The solid curve in (b)
is the fit assuming a flat background only. 
Both graphs include first pulses with measured energy of 50-500~keV 
and second pulses with energy greater than 100~keV.}
\label{datadelay}
\end{center}
\end{figure}

Data included in the analysis of neutrons from rock 
were collected during 123 days of live 
time. After rejection of events from these two sources of background 
there remained two types of events in the data: neutron events caused 
by radioactivity in rock and random coincidences of background pulses 
within a 200~$\mu$s time window. The average rate of events was about 
2.0/min, reduced down to about 10 events/day after cuts. 
The energy spectra of the first and second pulses,  
for all events included in the analysis, before and after the cuts
are shown in Figures~\ref{datasp_nocuts} and \ref{datasp} respectively. 
In addition, the plots of tail/total ratio of these events, 
before and after the cuts,
are shown in Figures~\ref{datatail_nocuts} and \ref{datatail} respectively. 
Figure~\ref{datadelay}a displays the distribution of time delays 
between the two pulses after cuts. 
Since neutron events are expected mainly at low 
energies, Figure \ref{datadelay}a displays only events between 50 
and 500 keV. Two populations are seen in Figure
\ref{datadelay}a (histogram with the best fit shown as a solid line) 
-- one is a flat distribution caused by 
random coincidences of pulses from background gammas and the other is 
an exponential distribution of time delay between the first pulse 
from proton recoils and the second pulse from neutron capture gammas. 
To obtain the number of neutron events, the measured distribution was 
fitted with the sum of two functions: flat distribution with unknown  
rate (free parameter) and an exponential with unknown number of 
neutron events (free parameter) and known index (from the neutron 
calibration runs) -- the same formula as used for the neutron calibration 
analysis. The result of the fit gave a total number of 
neutron events $224 \pm 79$ which corresponds to $1.82 \pm 0.64$ 
events per day, in the energy range of 50-500~keV. 
Allowing the index of the exponential to vary 
within bounds ($\pm 20\%$ of the value from calibration or 
approximately 2$\sigma$), a similar result was found.

A second long run was carried out for 3 months, with the same settings 
as for the neutron measurements. Neutron shielding 
(about 10 cm of polypropylene) was placed around the 'castle', 
except the bottom part. For this configuration only about 20\% of the 
rock neutrons were expected relative to the unshielded detector. The 
aim of this run was to show that the neutrons observed without neutron 
shielding are indeed coming from the rock. Figure \ref{datadelay}b shows 
the resulting time delay distribution for this run, a nearly flat
distribution, consistent with the detection of mainly 
random coincidences. This is different from the time 
delay distribution from the unshielded run (Figure \ref{datadelay}a).
From the fit to the time delay distribution, having fixed the mean 
time delay (from the neutron calibration runs), it was found that the
number of neutrons was $44 \pm 69$ (normalised to the exposure 
time of the neutron measurements), consistent with 0 and certainly much 
less than in the unshielded run.

To convert the measured rate into neutron flux, we need to know the 
neutron detection efficiency. To determine this efficiency,
a Monte Carlo simulation of neutron production, transport and 
detection can be used. Such a simulation, carried out for a certain
concentration of U/Th, allows us to calculate the neutron flux
above a certain threshold and the rate of proton recoils and delayed
coincidences. Normalising the calculated rate of delayed coincidences 
to the measured one, we can obtain then the true value for the neutron flux
and U/Th concentrations. Relying totally on the Monte Carlo, however,
would mean that our result is model dependent. To avoid this,
we used both Monte Carlo simulations and calibration with the neutron
source to determine the neutron detection efficiency.

\begin{figure} [h!]
\begin{center}
\includegraphics[scale=0.8]{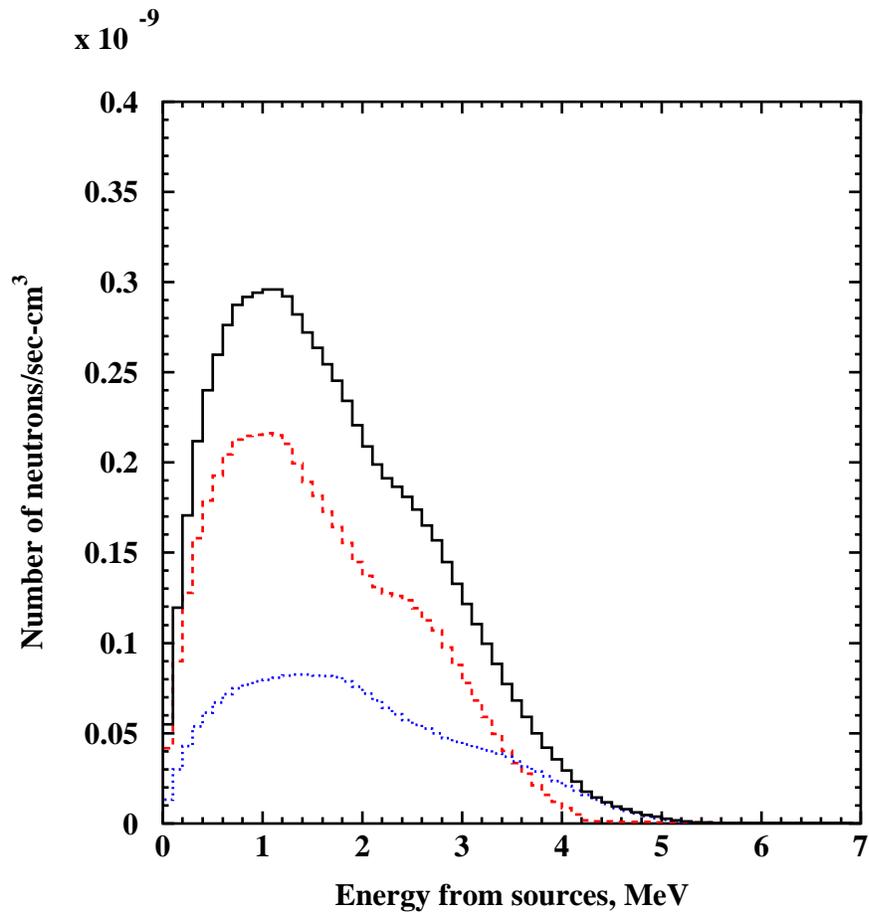}
\caption{Energy spectra of neutrons calculated with modified SOURCES. Spectra 
due to contamination of 10~ppb of U (dashed histogram, middle) and 
10~ppb of Th (dotted histogram, bottom) are shown. 
Solid histogram shows the total neutron production rate due 
to 10 ppb of U and Th in NaCl.}
\label{uthsources}
\end{center}
\end{figure}

The Monte Carlo simulations were carried out in several stages, 
using GEANT4. For the first stage, neutrons generated via spontaneous 
fission and ($\alpha$,n) reactions from $^{238}$U and $^{232}$Th
decay chains (calculated using the modified code 
SOURCES~\cite{sources,carson04}), were propagated to the rock-laboratory 
boundary and their parameters (such as energy, neutron position and 
the 3-momentum) were recorded. Figure \ref{uthsources} shows the 
spectra of neutrons from 10~ppb of $^{238}$U and 10~ppb of $^{232}$Th, 
calculated using SOURCES. These were the input spectra for the first 
stage of the GEANT4 simulations. At the second stage, these neutrons 
were propagated from the rock-laboratory boundary to the outer surface 
of the shielding (castle made of Cu and Fe). At the next stage, neutrons 
were transported to  the outer surface of the scintillator 
(neutron parameters were always recorded)
and, finally, inside the sensitive volume of the detector where they produced 
proton recoils by elastic scattering in the scintillator.
The same procedure was applied to neutrons from the $^{252}$Cf source, 
starting from neutron propagation from Cf source to the shielding.

As a result of the simulations the absolute rate and energy spectra of 
neutrons at different surfaces and those of proton recoils in the 
detector were obtained. The simulated proton recoil rate from $^{252}$Cf 
source placed at a distance of 1~m away from the detector (same position 
as in the calibration run) was 14.51~s$\ ^{-1}$ at 50-500~keV. Typical 
statistical error for all simulation results is
less than 1\%. This is much less than the experimental statistical
and systematic uncertainties and will be neglected further on.
The corresponding proton recoil rates from 10~ppb U and 10~ppb Th were 
found to be 4.22~d$^{-1}$ and 2.18~d$^{-1}$, respectively.

The efficiency of detecting delayed coincidences was calculated as 
the ratio of the measured rate (delayed coincidences with 100 keV 
threshold for the 2nd pulse) from Cf source to the calculated rate of  
single proton recoils at 50-500~keV and was found to be 0.037 $\pm$ 0.003. 
An efficiency of 0.036 $\pm$ 0.004 was obtained when the measured rates 
of delayed coincidences and the measured rate of single pulses (mainly 
proton recoils) were used, in agreement with the value obtained
from simulation of proton recoils. (Note that good agreement was 
found between the measured and simulated rates and spectra of proton 
recoils from the Cf source, as shown in Figure~\ref{rate_protonrec} and 
discussed in Section~\ref{calibrations}). The same efficiency was then used 
to convert the calculated rate of proton recoils from background neutrons 
into the expected rate of delayed coincidences.

\begin{figure} [h!]
\begin{center}
\includegraphics[scale=0.8]{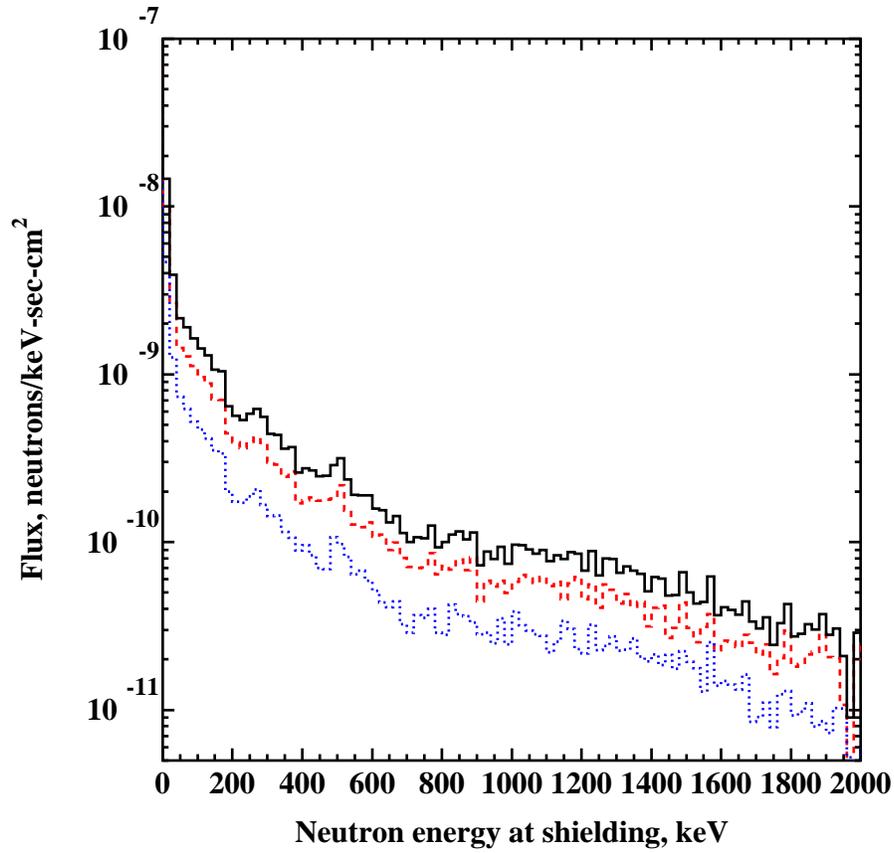}
\caption{Energy spectra of neutrons at the outer surface of the shielding 
due to contamination of 10~ppb U (dashed histogram, middle) and 10~ppb Th 
(dotted histogram, bottom), calculated using GEANT4. The total background 
(the sum of U and Th 
contributions) is shown with the solid histogram.}
\label{bgshielding}
\end{center}
\end{figure}

\begin{figure} [h!]
\begin{center}
\includegraphics[scale=0.8]{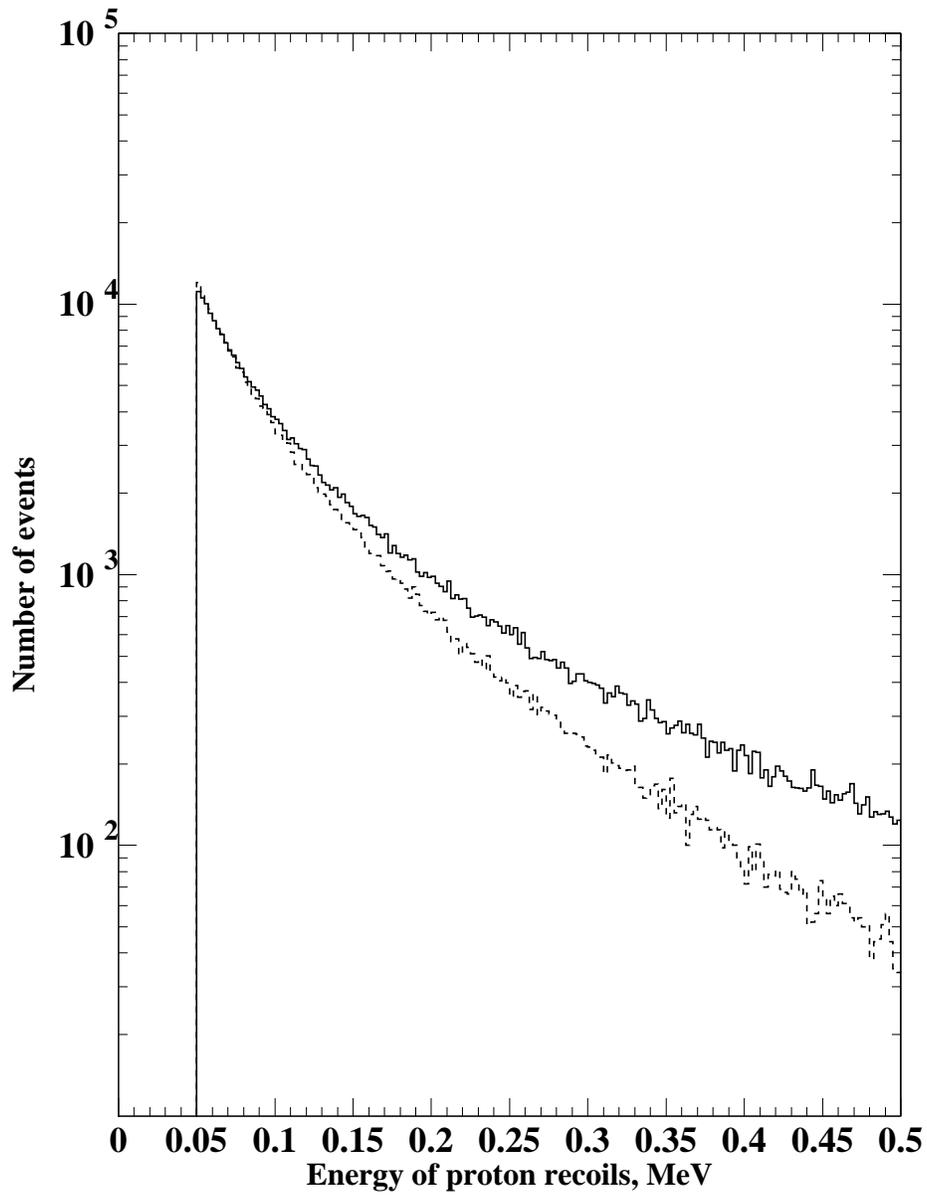}
\caption{Energy spectra of proton recoils due to background (U and Th, 
dashed histogram) and due to Cf neutrons (solid histogram), calculated 
using GEANT4. Cf spectrum was normalised to the background spectrum at 50 keV.}
\label{eprotonrec}
\end{center}
\end{figure}

\begin{figure} [h!]
\begin{center}
\includegraphics[scale=0.8]{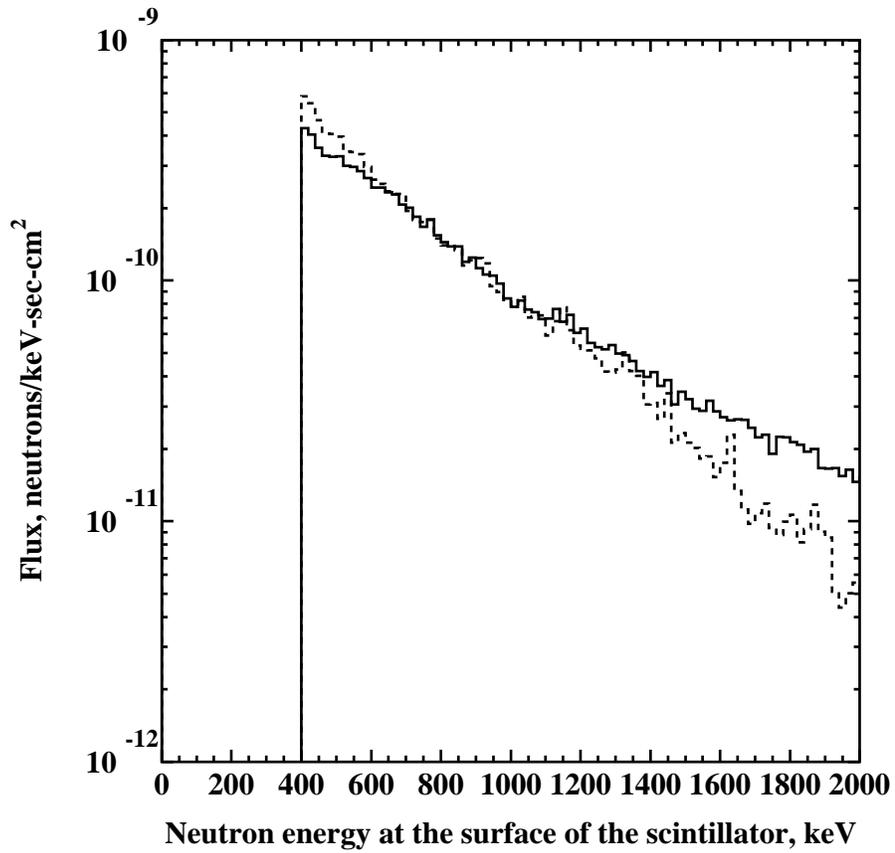}
\caption{Flux of neutrons, which produced proton recoils, from the 
background (dashed histogram) and the Cf source (solid histogram), at 
the outer surface of the scintillator, calculated using GEANT4. 
The spectrum of Cf was normalised to the integrated spectrum of the 
background (total fluxes above 0.4~MeV are assumed to be the same).}
\label{etarget_pr}
\end{center}
\end{figure}

The simulated fluxes of neutrons at the outer surface of the shielding, 
from 10~ppb U, 10~ppb Th and the sum, 
are shown in Figure~\ref{bgshielding}. The energy spectra of proton 
recoils from GEANT4 simulations of 
U/Th background (the sum of the contributions from the two 
isotopes) and $^{252}$Cf are compared in Figure~\ref{eprotonrec}.
The relative normalisation is done in order to have similar
flux at about 50 keV and to compare the shapes of the
two spectra. Figure~\ref{etarget_pr} depicts the simulated fluxes of neutrons
from the background and Cf source at the outer surface of the scintillator,
which produced the proton recoil spectra shown in Figure ~\ref{eprotonrec}.
The spectrum from Cf was normalised to the overall area of the spectrum 
from the background. The shapes of the two spectra are quite similar
with an estimated difference in total rate of 11\% in the energy range of 
0.4-1~MeV. Neutrons from this region contribute the most to the proton 
recoils at 50-500~keV
and also are more likely to be captured in the scintillator. 
We used this difference of 11\% as an estimate of systematic uncertainty
from this method of efficiency calculation (see Section~\ref{uncertainties}
for discussion on systematic uncertainties).

\section{Results}\label{uncertainties}

In this Section, the results from the above studies are presented.
In addition, all the efficiencies and uncertainties, which are taken 
into account in the final results, are discussed.

To investigate the effect of the dead time of DAQ on the rate of 
delayed coincidences in the background run, another short background 
run in which single pulses were recorded was carried out. From 
this run the rate of single pulses was obtained and the expected 
rate of random coincidences for the background run with two pulses 
recorded above 50 keV was calculated as $20.57 \pm 0.36$ events per day. 
This is in good agreement with the measured rate of $23.02 \pm 0.78$ 
events per day (the flat component in the time delay distribution 
similar to that shown in Figure~\ref{datadelay} but without 
cuts, except the energy threshold of 50 keV for both pulses in an event).  
This demonstrates that the dead time does not affect the 
background measurements. 

To evaluate the neutron flux, the trigger efficiency and the efficiencies 
of the various selection criteria should be taken into account.
The energy dependent trigger efficiency can be calculated 
from the energy spectrum of events from $^{252}$Cf (single events 
without delayed coincidences) taken with two different energy thresholds, 
the threshold used in the experiment and another one half of this (5~mV). 
The ratio of events for specific energy gives the energy dependent
trigger efficiency for the measurements with the higher threshold. It 
was found that the efficiency is 100\% for energies more than 50~keV, 
the spectra measured with the two thresholds being practically
identical above 50 keV. Note that the hardware energy threshold
was set at about 20 keV, 2 photoelectrons from each PMT.

As mentioned before, the ratio of the observed charge in the tail of 
the pulse to the total charge was used to reject the correlated 
background from Bi$\rightarrow$Po$\rightarrow$Pb decay.
The efficiency of this cut was calculated using data 
from a neutron calibration run with the $^{252}$Cf source 
with a software threshold of 50 keV for which the trigger efficiency 
is 100\%. The fraction of events from $^{252}$Cf which 
passed the cut above the software threshold of 50~keV yields an 
efficiency of (61 $\pm$ 1)\%. The value was obtained using fits to the time 
delay distributions similar to those shown in Figure~\ref{timedelay}.
Taking into account this cut efficiency, the efficiency of detecting 
delayed coincidences discussed in Section~\ref{measurements}
was reduced to 0.023$\pm$0.002.

Table~\ref{rates} shows a summary of all the measured and predicted
rates of proton recoils and coincidences from $^{252}$Cf and the 
backgound runs. Also the efficiency of detecting coincidences is shown,
taking into account the efficiency of the tail/total cut.

\begin{figure} [h!]
\begin{center}
\includegraphics[scale=0.8]{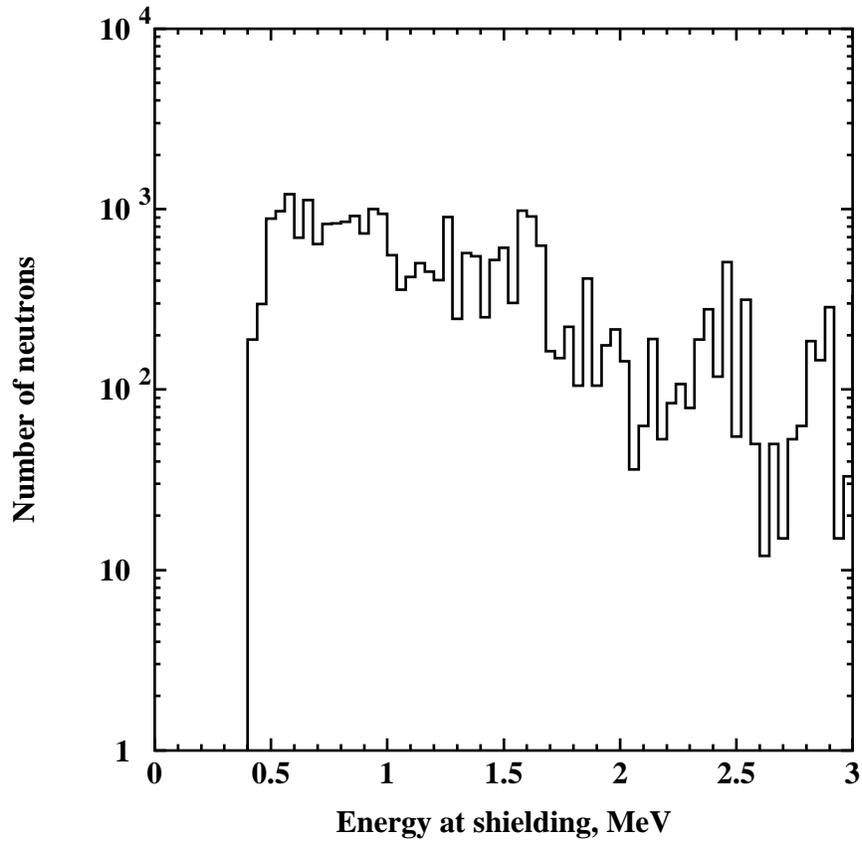}
\caption{Energy spectrum of neutrons (due to 100~ppb U and Th) at the outer 
surface of the shielding, which produce proton recoils with measured 
energy 50-500~keV. The spectrum was simulated using GEANT4. 
A threshold of 0.5~MeV for the measured neutron flux was estimated from 
this graph.}
\label{eshield}
\end{center}
\end{figure}

The simulated energy spectrum of neutrons at the outer surface of shielding 
which produced proton recoils in the detector (within 50-500~keV energy 
range), as shown in Figure~\ref{eshield}, indicates a threshold of 0.5~MeV for 
the measured neutron flux. The features seen in Figure~\ref{eshield} are
not statistical but reflect the structures in the cross-section of 
neutron scattering on Na in the rock. 
The threshold at the rock-laboratory boundary 
is the same, since for most neutrons recorded at the surface of the shielding
no interactions take place between the surface of the rock and the surface of 
the shielding.

Assuming 10~ppb U and 10~ppb Th, the neutron flux at the rock-laboratory 
boundary was calculated as $9.11 \times 10^{-8}$~cm$^{-2}$~s$^{-1}$ 
above 0.5~MeV. This flux corresponds to neutrons which enter 
the laboratory hall for the first time, so does not take into account further
scattering of neutrons from the walls. When this effect is included 
(neutrons which cross the cavern and scatter back from the rock into 
the cavern again), the flux at shielding is found to be 
$1.37 \times 10^{-7}$~cm$^{-2}$~s$^{-1}$ above 0.5~MeV. This is 51\% 
higher than the value at the rock-laboratory boundary, in agreement 
with Lemrani et al.~\cite{lemrani}, who claimed a 50\% increment to the 
flux above 1~MeV due to the contribution of the back-scattered neutrons.
For this flux a rate of 6.40~d$^{-1}$ for proton recoils at 50-500~keV was
obtained. Taking into account the efficiency of detecting delayed
coincidences of $0.023\pm 0.002$ we get a predicted rate of events in the
background run of $0.147\pm 0.013$ per day. As the detector in fact 
measured a rate of $1.82 \pm 0.64$ per day, the neutron flux 
was $(1.70\pm 0.60)\times 10^{-6}$~cm$^{-2}$~s$^{-1}$ above 0.5~MeV at
shielding, and $(1.13 \pm 0.40) \times 10^{-6}$~cm$^{-2}$~s$^{-1}$ above
0.5~MeV at the entrance to the cavern (no back scattering). 
An increase in the event rate of 1\% is predicted by the simulations
if proton recoils are considered without an upper energy cut of 0.5~MeV.
Correcting for this increase we obtain the final neutron flux
as $(1.72\pm 0.61)\times 10^{-6}$~cm$^{-2}$~s$^{-1}$ above 0.5~MeV at
shielding, and $(1.14 \pm 0.40) \times 10^{-6}$~cm$^{-2}$~s$^{-1}$ above
0.5~MeV at the entrance to the cavern.
This result was obtained assuming equal concentrations of U and Th
in rock. As the expected neutron spectra from U and Th after 
propagation in rock 
and shielding are very similar (see Figure~\ref{bgshielding}) and the 
spectra of proton recoils produced by these neutrons are also similar,
different assumptions about the relative fractions of U and Th result
in very similar values for neutron flux, the difference being of the order 
of a few percent.
Our measured neutron flux $(1.72\pm 0.61)\times 10^{-6}$~cm$^{-2}$~s$^{-1}$ 
above 0.5~MeV agrees well with the simulation result 
$1.21 \times 10^{-6}$~cm$^{-2}$~s$^{-1}$ assuming measured concentrations
of 67 ppb of U and 127 ppb of Th \cite{peter}.

The same re-scaling, as was used to derive the neutron flux, should 
be applied to the original concentrations of U and Th
resulting in $127 \pm 45$~ppb of U and Th assuming their equal
concentrations. If we assume that Th is twice as abundant as U,
as measured by Smith et al. \cite{peter}, then the resulting
concentrations are $95 \pm 34$~ppb of U and 
$190 \pm 69$~ppb of Th. These values agree within errors with
the recent measurements of the U and Th concentrations at Boulby
\cite{peter}. They are also consistent 
with most of the previous measurements of the U and Th concentrations in 
Boulby rock using different techniques keeping in mind the large spread 
of the measured values \cite{ukdmc}. Table~\ref{flux} summarises the
calculated and measured fluxes of neutrons at different surfaces, with 
different concentrations of U and Th.

The errors given above are purely statistical, determined by the fit 
to the measured time delay distribution. In addition to the 
statistical errors we can estimate systematic uncertainties.
These are associated with the following parameters used to evaluate the 
neutron flux:
(i) conversion of the charge to the energy scale for proton recoils;
(ii) estimation of efficiencies;
(iii) accuracy of simulations.

The uncertainty in the measured energy of proton recoils has
two contributions: possible error and stability of the energy 
calibration using $^{60}$Co and $^{137}$Cs sources 
and uncertainty in the quenching factor for proton recoils. 

A 5\% change in the charge-to-energy conversion factor or in the width 
(standard deviation or $\sigma$) of the gaussian function for the 
energy resolution, changes significantly the calculated spectra 
of events from gamma sources, so the agreement with measurements
seen in Figure~\ref{encal} disappears. The uncertainty in the quenching 
factor for proton recoils is 
more difficult to estimate. There is, however, a way to estimate the 
possible error due to the two factors together (charge-to-energy conversion 
and quenching factor).
As was mentioned before, in Figure \ref{corrsp}, we observe an alpha 
peak in the correlated background at 
0.8 MeV, but according to the quenching of alphas \cite{verbinski} the peak 
should be at about 0.96 MeV. This 20\% difference in the peak 
position can be used as an estimate of the uncertainty in the 
energy scale for proton recoils. Changing the software energy 
threshold by 20\% from 50 keV to 60 keV results in a decrease of 
the measured rate from $^{252}$Cf giving a possible error in the flux 
intensity of 10\%. This exceeds significantly the estimated uncertainty
in the charge-to-energy conversion factor and energy resolution (5\%). 

One of the uncertainties in the efficiency estimate comes from the
statistical error of the fit to the time delay distribution for the 
calibration run with the Cf source. This error is of the order of 8\%.
Another uncertainty in the efficiency comes from
the difference between the shape of the flux of neutrons from
the $^{252}$Cf source and background, at the outer 
surface of the scintillator, as shown in Figure~\ref{etarget_pr}, which was 
calculated in Section~\ref{measurements} as 11\%.

The simulations directly affect the neutron flux evaluated from the 
measured rate through the charge-to-energy conversion, proton recoil 
detection efficiency (including quenching), simulated energy resolution
and other effects. In fact the accuracy of simulations has partly been 
considered when the corresponding effects have been studied (see above). 
The measured and predicted proton recoil rates from the neutron source 
agree within 14\%. The small difference arises from the presence of 
gammas from neutron captures in the data and their absence in the 
simulations. We estimate then the uncertainty due to simulations as 14\%.
The total systematic uncertainty can be found by adding the four 
errors discussed above in quadrature resulting in a value of 22\%.
A summary of all the systematic uncertainties is shown in 
Table~\ref{systuncert}.

\section{Conclusions}

A technique to measure low intensity neutron flux has been 
developed based on a detection of delayed coincidences
between proton recoils and neutron capture gammas.
The neutron flux in the Boulby Underground Laboratory at a depth of 
2800~m~w.e. was measured using liquid scintillator detector 
as 
$(1.72 \pm 0.61 (stat.) \pm 0.38 (syst.)) \times 10^{-6}$~cm$^{-2}$~s$^{-1}$ 
above 0.5~MeV including the effect of neutron back-scattering
from the cavern walls.
This flux corresponds to the concentrations of 
either$\ 127 \pm 45 (stat.) \pm 28 (syst.)$~ppb 
of U and Th, assuming their concentrations are equal or 
$95 \pm 34 (stat.) \pm 21 (syst.)$~ppb of U and
$190 \pm 69 (stat.) \pm 42 (syst.)$~ppb of Th, 
if Th is twice as abundant as U.

\section{Acknowledgments}
This work has been supported by the ILIAS integrating activity
(Contract No. RII3-CT-2004-506222) as part of the EU FP6 programme 
in Astroparticle Physics. We aknowledge the financial support from 
the Particle Physics and Astronomy Research Council (PPARC). 
We also like to thank the Cleveland Potash Ltd for assistance.

\clearpage
\begin{table}[h!]
\begin{center}
\caption{Measured and predicted rates of single proton recoils and 
delayed coincidences from the $^{252}$Cf and the backgound runs. 
Rows 1 and 2 show the calculated and the measured rate of single proton 
recoils without delayed coincidences from $^{252}$Cf source on top of 
the castle. In rows 3 and 4, the measured rate of delayed coincidences 
and the efficiency of detecting delayed coincidences, 
calculated from the measured values, are shown respectively. In rows 5-7
similar rates and efficiency are shown for $^{252}$Cf source at a distance 
of 1~m away from the detector. No run with single proton recoil 
measurements, with a source at a distance of 1~m from the detector, 
was performed before the neutron shielding was installed 
around the detector.
In this case the efficiency is calculated from 
the calculated rate of single proton recoils and the measured rate of 
delayed coincidences. The last row shows the measured rate of delayed 
coincidences for the background run. In the efficiency calculation, 
the efficiency of the tail/total cut was taken into account (see 
Section~\ref{uncertainties}).}
\begin{tabular}{|l | l | c|}\hline
& Calculated rate of single proton recoils, s$^{-1}$ & 140.30 \\
$^{252}$Cf source on top & Measured rate of single proton recoils, s$^{-1}$ & 163.88 $\pm$ 1.40\\
of the castle & Measured rate of delayed coincidences, s$^{-1}$ & 5.87 $\pm$ 0.68\\
& Efficiency of detecting delayed coincidences & 0.022 $\pm$ 0.002\\
\hline
$^{252}$Cf source 1~m away & Calculated rate of single proton recoils, s$^{-1}$ & 14.51\\
from the detector & Measured rate of delayed coincidences, s$^{-1}$ & 0.53 $\pm$ 0.04\\
& Efficiency of detecting delayed coincidences & 0.023 $\pm$ 0.002\\
\hline
Background run & Measured rate of delayed coincidences, d$^{-1}$ & 1.82 $\pm$ 0.64\\
\hline
\end {tabular}
\label{rates}
\end{center}
\end{table}

\clearpage
\begin{table}[h!]
\begin{center}
\caption{Flux of neutrons at different surfaces with different concentrations 
of U and Th. The first column shows the concentrations of U and Th in rock.
At the beginning,  they were assumed to be 10~ppb U and 10~ppb Th (rows 1-4). 
Later, the real concentrations of U and Th were found from the data analysis 
(rows 5-6), assuming equal concentration of U and Th.
The second column shows the surface on which
the flux was calculated, for different concentration of U, Th and the 
last column shows the corresponding fluxes.
The errors given for the measured concentrations and fluxes are statistical.}
\vspace{0.5cm}
\begin{tabular}{|l |l| c|}\hline
Concentration in rock & Surface & Flux above 0.5~MeV\\ [0.5ex]
& & [cm$^{-2}$~s$^{-1}$]\\
\hline 
& & Simulated Flux\\
10ppb U  & Rock w/o back scattering & $6.02 \times 10^{-8}$\\
10ppb Th & Rock w/o back scattering  & $3.09 \times 10^{-8}$\\
10ppb U & Outer surface of shielding & $ 9.06\times 10^{-8}$\\
10ppb Th & Outer surface of shielding & $4.66 \times 10^{-8}$\\
\hline
& & Measured Flux\\
(127 $\pm$ 45) ppb U and Th & Rock  w/o back scattering & $(1.14 \pm 0.40) \times 10^{-6}$\\
(127 $\pm$ 45) ppb U and Th & Outer surface of shielding & $(1.72\pm 0.61)\times 10^{-6}$\\
\hline
\end {tabular}
\label{flux}
\end{center}
\end{table}

\clearpage
\begin{table}[h!]
\begin{center}
\caption{Systematic uncertainties.}
\begin{tabular}{|l | c|}\hline
Sources of systematic uncertainties & Relative\\[0.5ex]
& systematic\\
& uncertainties\\
\hline 
Charge to energy conversion and quenching factor & 0.10\\
Fit to the time delay distribution for $^{252}$Cf run & 0.08\\
Difference between neutron spectra from $^{252}$Cf source and background & 0.11\\
Difference between measured and simulated recoil rates & 0.14\\
\hline
Total systematic uncertainty & 0.22\\
\hline
\end {tabular}
\label{systuncert}
\end{center}
\end{table}

\clearpage

\clearpage

\clearpage

\clearpage

\clearpage

\clearpage

\clearpage

\clearpage

\clearpage

\clearpage

\clearpage

\clearpage

\clearpage

\clearpage

\clearpage
 
\clearpage

\clearpage

\clearpage

\clearpage

\clearpage
 
\end {document}